\documentclass[10pt,prd,twocolumn,eqsecnum,amssymb,amsmath,showpacs,a4paper,superscriptaddress,nofootinbib,floatfix]{revtex4-2}
\usepackage{import}

\usepackage{graphicx}
\usepackage{amsfonts}
\usepackage{amsmath}
\usepackage{hyperref}
\usepackage{units}
\usepackage{color}
\usepackage{dcolumn}
\usepackage{bm} 
\usepackage{cleveref}
\usepackage{braket}
\usepackage{todonotes}
\crefname{section}{§}{§§}
\Crefname{section}{§}{§§}
\usepackage{mathtools}
\usepackage{chngcntr}
\counterwithout{equation}{section}
\usepackage[normalem]{ulem}
\usepackage{tikz}
\usepackage{comment}
\usepackage{nicefrac}
\usepackage{ragged2e}

\usepackage[T1]{fontenc}
\usepackage[utf8,applemac]{inputenc}

\newcommand{\be}{\begin{equation}}
\newcommand{\ee}{\end{equation}}
\newcommand{\bsub}{\begin{subequations}}
\newcommand{\esub}{\end{subequations}}
\newcommand{\bea}{\begin{eqnarray}}
\newcommand{\eea}{\end{eqnarray}}
\newcommand{\bi} {\begin{itemize}}
\newcommand{\ei} {\end{itemize}}
\newcommand{\bmat} {\begin{pmatrix}}
\newcommand{\emat} {\end{pmatrix}} 

\usepackage[final]{showlabels} 

\newcommand{\D}{\mathrm{d}}
\newcommand{\I}{\mathrm{i}}
\newcommand{\E}{\mathrm{e}}

\newcommand{\mrm}[1]{\mathrm{#1}}

\DeclareMathOperator{\sgn}{sgn}
\DeclareMathOperator{\Realpart}{Re}

\makeatletter
\newcommand*{\balancecolsandclearpage}{%
  \close@column@grid
  \clearpage
  \twocolumngrid
}
\makeatother

\makeatletter
\let\cat@comma@active\@empty
\makeatother

\usepackage{xcolor}
\hypersetup{  colorlinks,
              linktoc         = page, 
              urlcolor        = {green!80!blue},
              linkcolor       = {orange!60!black}, 
              urlcolor        = {cyan!80!black}, 
              citecolor       = {cyan!80!black}, 
              anchorcolor     = {yellow}
}

\begin{document}


\title{Nondestructive optomechanical detection scheme for Bose-Einstein condensates}

\author{Cisco Gooding}
\affiliation{Laboratoire Kastler Brossel, Sorbonne Universit\'e, ENS-Universit\'e PSL, CNRS, Coll\`ege de France, 4 place Jussieu, Paris F-75252, France}
\author{Cameron R. D. Bunney}
\affiliation{School of Physics and Astronomy, University of Nottingham, University Park, Nottingham, NG7 2RD, United Kingdom}
\author{Samin Tajik}
\affiliation{Department of Physics \& Astronomy, University of British Columbia, Vancouver, Canada V6T 1Z1}
\author{Sebastian Erne}
\affiliation{Vienna Center for Quantum Science and Technology (VCQ), Atominstitut, TU Wien, Stadionallee 2, 1020 Vienna, Austria}
\author{Steffen Biermann}
\affiliation{School of Mathematical Sciences, University of Nottingham, University Park, Nottingham, NG7 2RD, United Kingdom}
\author{J\"org Schmiedmayer}
\affiliation{Vienna Center for Quantum Science and Technology (VCQ), Atominstitut, TU Wien, Stadionallee 2, 1020 Vienna, Austria}
\author{Jorma Louko}
\affiliation{School of Mathematical Sciences, University of Nottingham, University Park, Nottingham, NG7 2RD, United Kingdom}
\author{William G. Unruh}
\affiliation{Department of Physics \& Astronomy, University of British Columbia, Vancouver, Canada V6T 1Z1}
\affiliation{Hagler IAS, IQSE, Texas A\&M, College Station, Texas, 77843-4242, USA}
\author{Silke Weinfurtner}
\affiliation{School of Mathematical Sciences, University of Nottingham, University Park, Nottingham, NG7 2RD, United Kingdom}
\affiliation{Centre for the Mathematics and Theoretical Physics of Quantum Non-Equilibrium Systems, University of Nottingham, Nottingham, NG7 2RD, United Kingdom}

\date{August 2025; revised December 2025.\\ aaPublished in Phys.\ Rev.\ Lett.\ \textbf{136}, 043401 (2026), doi.org/10.1103/4yfh-tm4f}

\begin{abstract}
We present a two-tone heterodyne optical readout scheme to extract unequal-time density correlations along an arbitrary stationary interaction path from a pancake-shaped Bose-Einstein condensate, using a modulated laser probe. Analysing the measurement noise both from imprecision and backaction, we identify the standard quantum limit for the signal-extraction scheme, and examine how a class of two-mode squeezed initial states can be used to push beyond this limit. As an application, we show how the readout scheme can be used for an experimentally feasible realisation of acceleration-dependence of quantum-vacuum fluctuations in the system, including the analogue spacetime circular motion Unruh effect. 
The scheme is adaptable beyond Bose-Einstein condensates, providing nondestructive access to unequal-time correlations in quantum fluids. 
\end{abstract}

\maketitle

\setlength{\belowdisplayskip}{1.5pt} \setlength{\belowdisplayshortskip}{1.5pt}
\setlength{\abovedisplayskip}{1.5pt} \setlength{\abovedisplayshortskip}{1.5pt}

\textbf{Introduction}---Bose-Einstein 
condensates (BECs) provide a powerful platform for exploring a range of intriguing quantum field theory processes, including information theoretical~\cite{Tajik2023_AreaLaw} and far-from-equilibrium phenomena such as quantum turbulence~\cite{Dogra2023,PhysRevLett.104.075301,PhysRevLett.103.045301,PhysRevLett.95.145301}, first-order relativistic phase transitions~\cite{Opanchuk2013,
  Fialko2015,
  Fialko2017,
  Braden2018,
  Billam2019PRL,
  Braden2019,
  Ng2021,
  Jenkins2024a,
  Zenesini2024,
  Cominotti2025}, 
and curved spacetime dynamics that mimic the early universe~\cite{UweAndRalf2004, Jaskula2012,Viermann2022, Tajik2023} and black holes~\cite{PhysRevLett.85.4643,PhysRevLett.105.240401,Steinhauer2016}. A key technical challenge lies in controlling the condensate, leading to innovations such as optical box traps~\cite{Navon2021, Gaunt2013} and multicomponent condensates~\cite{Jenkins2024b}. A common objective across these developments is the advancement of novel detection schemes. 

Currently, state-of-the-art detection methods are primarily destructive, capturing wave dynamics frozen in time~\cite{Schweigler2017,Tajik2023_AreaLaw}. Although this approach has been highly successful, enabling the extraction of statistical correlations and cumulants of density fluctuations between different points in the condensate, there remains strong motivation to develop improved, potentially nondestructive detection techniques~\cite{PhysRevA.67.043609}. A specific candidate technique was introduced in~\cite{PhysRevLett.125.213603}, where a focused laser beam passes perpendicularly through a pancake-shaped BEC, sampling density fluctuations locally over time. The point of interaction between the probe beam and the condensate need not be static but can be placed in controlled motion. The probe beam thus records the local density fluctuations encountered along the interaction path, encoding the unequal space and time correlations and their dependence on the motion~\cite{PhysRevLett.125.213603,UweAndPetrObsDep2004}.

The measurement sensitivity in this detection technique is set by the competition between shot noise in the probe and radiation-pressure backaction onto the condensate. 
In the present Letter we go significantly beyond \cite{PhysRevLett.125.213603} by 
providing a complete end-to-end detection theory that controls the sensitivity: we derive the standard quantum limit (SQL) \cite{caves-PRL,second} for a two-tone heterodyne readout of a moving-point optical probe, include the full nonlinear Bogoliubov backaction structure, and show how two-mode squeezed input states enable sub-SQL sensitivity while preserving nondestructiveness. The scheme is adaptable beyond BEC, providing non-destructive access to unequal-time correlations in quantum fluids. 

As an application, we apply our interferometric machinery to accelerated paths of the interaction point. We detail an experimentally feasible detection and readout scheme to extract unequal-time BEC correlations from the beam. We show that these correlations bear an imprint of the acceleration of the interaction-point path. This acceleration dependence renders the two-dimensional BEC a powerful platform for the quantum simulation of acceleration effects in quantum field theory, particularly the celebrated Unruh effect~\cite{Fulling,Davies1975,unruh76,Fulling:2014}. \\

\textbf{Phonon detectors---}We describe a BEC of atomic mass $m$, confined to a plane (defined as $z=0$) by a trapping potential. Upon integration in $z$, the BEC is described by a field $\Phi(t,\bm{x})$ governed by the two-dimensional Gross-Pitaevskii Lagrangian,
\begin{align}
    L_{\text{BEC}}~=~\int \D^2 \bm{x} \left[\I\Phi^\dagger \partial_t \Phi-\frac{1}{2m}|\nabla\Phi|^2+\frac{g_{2d}}{2}|\Phi|^4\right]\,,
    \label{L_BEC}
\end{align}where $\bm{x}=(x,y)$ and $g_{2d}$ is the $s$-wave scattering strength. When we consider the real part of perturbations about a background condensate $\Phi_0$, these long-wavelength phonons evolve as a massless Klein-Gordon scalar field~\cite{PhysRevLett.85.4643,Unruh:2022gso}. In terms of the physical BEC parameters, this corresponds to density fluctuations about a background density $\rho_0=|\Phi_0|^2$, as detailed in Supplemental Material~\cite{suppmat}. 

\begin{figure}
    \centering
    \includegraphics[width=\columnwidth]{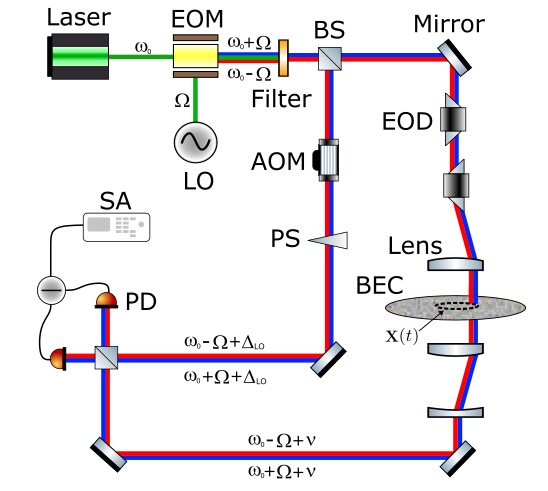}
    \caption{Optical circuit diagram for sampling BEC density fluctuations along a stationary interaction point trajectory, using a modulated laser probe with two sidebands. The band centred at $\omega_+ = \omega_0+\Omega$ ($\omega_- = \omega_0-\Omega$) is shown in blue (red). Post-interaction, the modulated signal is heterodyned using a two-tone reference beam, in which an acousto-optic modulator has inserted a frequency shift~$\Delta_{\mrm{LO}}$. 
    EOM: electro-optic modulator; BEC: Bose-Einstein condensate; BS: beamsplitter; EOD: electro-optic deflector; LO: local oscillator; AOM: acousto-optic modulator; PS: phase-shifter; PD: photodiode; SA: spectrum analyser.}
    \label{fig:exp}
\end{figure}

Long-wavelength phonons may be probed with a highly focused laser propagating transversely through the condensate parallel to the $z$ direction; the effective refractive index of the BEC shifts the laser phase. As shown in the optical circuit diagram of Figure~\ref{fig:exp}, the beam probing the condensate is prepared by sending monochromatic laser light of frequency $\omega_0$ into an electro-optic modulator (EOM), and then filtering out the central frequency band. The resulting laser field $E(t)$ has two prominent modulation bands peaked at $\omega_\pm\coloneqq \omega_0\pm\Omega$ (with $\Omega\ll\omega_0$), which may be expressed as
\begin{subequations}\label{eqn:: plus minus decomp}
\begin{align}
    E(t)~=~&E_0(\omega_0)(a(t)+a^\dagger(t))\,,\\
    a(t)~=~&a_+(t)+a_-(t)\,,\\
    a_\pm(t)~=~&(\alpha+\delta a_\pm(t))\E^{-\I\omega_\pm t}\,,
\end{align}    
\end{subequations}
where $E_0(\omega_0)$ is a dimensionless real-valued prefactor evaluated at the central frequency~$\omega_0$, $a_\pm$ are the mode operators for each modulation band, the full mode operator $a$ satisfies $[a(t),a^\dagger(t')]=\delta(t-t')$, and, without loss of generality, the coherent amplitude $\alpha$ is assumed to be real. We note that our notation suppresses spatial dependence for brevity, despite the laser serving as a local probe.

The interaction between the laser and the BEC can be modelled as the interaction between a two-dimensional scalar field $\phi(t,\bm{x})$---representing the density fluctuations in the BEC---and a one-dimensional scalar field $E(t)$---representing the laser---that is localised on the path $\bm{x}=\bm{X}(t)$. By arranging $\omega_0$ to coincide with an atomic resonance in the BEC, the phase shifts in the two modulation bands will be equal and opposite, as represented by the mode-operator transformation
\begin{equation}\label{eqn:: a transformation}
    a_\pm(t)~\rightarrow~\E^{-\I(\omega_\pm t\pm\psi_0)}\left(\alpha+\delta\tilde{a}_\pm(t)\right)\,,
\end{equation}
where $\psi_0$ is the time-independent phase shift due to the BEC bulk density. A tilde above an operator denotes that operator post interaction. These equal and oppositely detuned sidebands cancel out the zeroth-order Stark potential, reducing disturbance to the BEC due to stirring~\cite{PhysRevLett.125.213603}.

The first-order interaction $\delta\tilde{a}_\pm(t)$ in~\eqref{eqn:: a transformation} is given by
\begin{equation}\label{eq:firstorder}
    \delta\tilde{a}_\pm(t)~=~\delta a_\pm(t)\mp\frac{1}{2}\I\varepsilon\alpha\phi(t)\,,
\end{equation}
where the $\mp$ sign originates from the opposite phase shifts, $\phi(t)\coloneqq \phi(t,\bm{X}(t))$ is the field evaluated along the interaction path $\bm{X}(t)$, and, for sidebands detuned far from resonance, the interaction coupling is given by $\varepsilon=2|\hat{\alpha}_R|\omega_0 \sqrt{m\rho_0}$, where $\hat{\alpha}_R$ is the real part of the atomic polarisability (see Appendix~A). Physically, the field $\phi(t)$ corresponds to density fluctuations in the BEC sampled along the trajectory traced by the laser as it intersects the BEC\null.

The typical non-dispersive frequency range of phonons in the BEC lies well below the modulation frequency~$\Omega$. We denote the boundary of this non-dispersive range by $\Delta$ and write the field as
\begin{align}\label{eqn:: phi decomp}
\phi(t)~=~\int_{-\Delta}^\Delta \frac{\D\nu}{2\pi}\E^{-\I\nu t}D_\nu\,,
\end{align}
where $0<\Delta\ll\Omega$, 
and $D_\nu$ are the annihilation (for $\nu>0$) and creation (for $\nu<0$) operators defined with respect to the frame comoving along the interaction path, such that $D_{-\nu}=D_\nu^\dagger$. 
For typical quasi-2D condensates, $\Delta/2\pi$ lies in the few-kHz range.
We write the pre-interaction and post-interaction operators $\delta a_\pm(t)$ and $\delta {\tilde a}_\pm(t)$ similarly 
in terms of their Fourier transforms as 
$\delta a_\pm(t)=(2\pi)^{-1}\int_{-\Delta}^{\Delta}
\D\nu\,\E^{-\I\nu t} \delta a_\pm[\nu]$ 
and 
$\delta {\tilde a}_\pm(t)=(2\pi)^{-1}\int_{-\Delta}^{\Delta}
\D\nu\,\E^{-\I\nu t} \delta {\tilde a}_\pm[\nu]$. 

Now, the laser-BEC interaction \eqref{eq:firstorder}
can be described in terms of the frequency-space operators as a Bogoliubov transformation, and this transformation can be completed nonperturbatively to contain the backaction on the laser probe, quadratic in the coupling parameter $\varepsilon$~\cite{Unruh:2022gso}. As shown in Appendix~B, the outcome is 
\begin{equation}\label{delta_atilde}
\delta\tilde{a}_\pm[\nu]~=~\delta a_\pm[\nu]\pm\frac{\I \mu D_\nu}{\sqrt{2}} \pm \frac{\mu^2}{4} \sgn(\nu)\delta a_{b}[\nu]\,,
\end{equation}
where $\mu\coloneqq -\varepsilon \alpha/\sqrt{2}<0$ is a dimensionless coupling parameter and $\delta a_{b}$ is the contribution from backaction, given by 
\begin{multline}\label{eq:baa}
    \delta a_{b}[\nu]=\delta a_-[\nu]+\delta a_-[-\nu]^\dagger-\delta a_+[\nu]-\delta a_+[-\nu]^\dagger\, .
\end{multline} 
The signum function in \eqref{delta_atilde} implies that backaction on the two sides of the modulation bands takes opposite signs. This may seem surprising; however, similar asymmetric backaction has been observed within optomechanical systems~\cite{Khalili2012}.

The term involving the BEC mode operator $D_\nu$ in \eqref{delta_atilde} is the signal. This shows that the beam acts as a phonon detector. The last term in~\eqref{delta_atilde}, representing backaction, shows that the laser also records the noise that it has injected into the BEC\null. 
The task of the detection scheme will be to separate the signal from the noise.

\textbf{Detection scheme---}We shall now present a detection scheme where the signal $D_\nu$ can be extracted from the post-interaction laser beam \eqref{delta_atilde} despite the backaction noise contained in the last term 
and the shot noise contained in the first term.
We assume from now on that the interaction point trajectory $\bm{X}(t)$ is stationary (such as in uniform circular motion), and that the initial state of the BEC is stationary (such as a vacuum state or a thermal state). 
The quantity we wish to extract is the Fourier transform of the BEC unequal-time density two-point function along the path of the laser-BEC interaction point, given by 
\begin{align}\label{eqn:: PSD}
S_{\phi \phi}[\nu]~=~\int \D t\, \E^{-\I\nu t}\braket{\phi(t)\phi(0)}\,. 
\end{align}Because of the stationarity, 
$S_{\phi \phi}[\nu]$ \eqref{eqn:: PSD} is the BEC power spectral density (PSD). 
The interest of $S_{\phi \phi}[\nu]$ is that it is a multiple of the transition rate of a pointlike two-state quantum system that moves along the path of the laser-BEC interaction point, had we such a quantum system at hand; $\nu$~is the system's energy gap, with $\nu>0$ for excitations and $\nu<0$ for deexcitations \cite{unruh76,DeWitt1979,PhysRevLett.125.213603,PhysRevD.102.085006}. 
The optical detection scheme hence must be able to distinguish between positive and negative values of~$\nu$. 

Inspired by optical demonstrations of mechanical sideband asymmetry~\cite{bowenmilburn}, 
we consider the two-tone heterodyne scheme depicted in Figure~\ref{fig:exp}. 
Post-interaction, the signal riding on modulation bands centred at $\omega_+ = \omega_0+\Omega$ and $\omega_- = \omega_0-\Omega$ is mixed with a reference beam whose frequency has been detuned by the amount~$\Delta_{\mrm{LO}}$, 
in the intermediate range $\Delta \ll \Delta_{\mrm{LO}} \ll 2\Omega$, using an acousto-optic modulator as shown in Figure~\ref{fig:exp}. 
The beamsplitter outputs are then converted into photocurrents via photodiodes and their difference is taken, yielding the difference-photocurrent~$i(t)$.

The observed difference-photocurrent $i(t)$ is sent to a spectrum analyser. The PSD $S_{ii}$ produced by the spectrum analyser is related to $S_{\phi\phi}[\nu]$
\eqref{eqn:: PSD} by 
\begin{align}\label{diffPSDinf0}
S_{ii}[\Delta_{\mrm{LO}}-\nu]=     \mu^2\left(S_{\phi\phi}[\nu]+\mathcal{N}[\nu;\mu^2]\right)\, ,
\end{align}
where 
\begin{align}\label{diffPSDadd0}
   \mathcal{N}[\nu;\mu^2]=\frac{1}{\mu^2}+\frac{3\mu^2}{8}+\sgn(\nu)\, ,
\end{align}
as we show in Supplemental Material~\cite{suppmat}, building on Glauber's theory of photodetection \cite{glauber63}
and the relation between photocurrent PSDs and electromagnetic correlators~\cite{Ou1987}. 
A measurement of $S_{ii}$ hence yields $S_{\phi\phi}[\nu]$ over the full frequency band $\nu \in (-\Delta,\Delta)$. 

In practice, two additional noise sources appear in the photocurrent: (i) relative intensity noise (RIN) of the probe and reference beams, and (ii) phase noise of the local oscillator. RIN enters primarily through common-mode fluctuations and is strongly suppressed by the balanced detection configuration shown in Figure~\ref{fig:exp} (typically by $30$-$40$ dB \cite{Yuen1983,quantumnoise1985,Collett1987}). LO phase noise contributes an additive term to $S_{ii}[\omega]$ centred at the intermediate frequency~$\Delta_{\mathrm{LO}}$; its contribution scales inversely with the reference beam power, and is minimised by operating the LO at high power. For commercially available low-phase-noise lasers these terms remain well below the SQL noise floor across the detection bandwidth.

\textbf{Noise minimisation---}The remaining task is to minimise the added noise term $\mathcal{N}[\nu;\mu^2]$ \eqref{diffPSDadd0} that contributes to the measured PSD $S_{ii}$ \eqref{diffPSDinf0} on par with the signal $S_{\phi\phi}[\nu]$ that we wish to infer. 
The first term in $\mathcal{N}[\nu;\mu^2]$ \eqref{diffPSDadd0} arises from shot noise, which represents sensor imprecision~\cite{Khalili2012}. 
The second term arises from the backaction found in \eqref{delta_atilde} and~\eqref{eq:baa}. 
The balance between the two terms depends on the parameter~$\mu$, which is proportional to the probe beam's coherent amplitude and can be tuned by adjusting the beam's intensity. 
The minimum value of the noise is attained at $\mu^2=2\sqrt{2/3}$, where the coherent amplitude satisfies 
$\alpha^2 = \alpha_{\mathrm{SQL}}^2
:=
4\,\sqrt{2}/(\sqrt{3}\varepsilon^2)$.
This minimum is known as the standard quantum limit (SQL), which is the optimal balance between the shot-noise-dominated regime ($\alpha^2 < \alpha^2_{\mathrm{SQL}}$) and the backaction-noise-dominated regime ($\alpha^2 > \alpha^2_{\mathrm{SQL}}$). 
The added noise at the SQL is 
\begin{align}\label{diffPSDaddSQL}
    \mathcal{N}_{SQL}[\nu]~=~\sqrt{\frac{3}{2
    }}+\sgn(\nu)\, .
\end{align}

To reduce the added noise \emph{below\/} the SQL value~\eqref{diffPSDaddSQL}, 
we replace the laser field's initial state by a two-mode squeezed state, with the real-valued squeezing parameter~$\lambda$, where $\lambda=0$ is the unsqueezed initial state considered above. 
Formula \eqref{diffPSDadd0} is then replaced by 
\begin{equation}\label{diffPSDadd}
    \mathcal{N}[\nu;\mu,\lambda]=\frac{1+\frac{3}{2}\sinh^2 \! \lambda}{\mu^2}+\frac{3\mu^2\, \E^{-2\lambda}}{8}+\E^{-\lambda}\cosh\lambda\,\text{sgn}(\nu)\,,
\end{equation}
as shown in Supplemental Material~\cite{suppmat}. 
Related demonstrations of squeezing-enhanced heterodyne or sideband detection include Li \emph{et al}.~\cite{PhysRevLett.82.5225} and references therein. Comparison with \eqref{diffPSDadd0}
shows that the squeezing has increased the shot noise, for either sign of~$\lambda$, while the backaction noise has increased for $\lambda<0$ but decreased for $\lambda>0$. For given~$\lambda$, the minimum value of the added noise is attained at 
\begin{align}\label{mulambda}
\mu^2 
= \mu_{\lambda}^2 
:=
2\,\mrm{e}^{\lambda}\,\sqrt{\frac{2}{3}+\sinh^2 \! \lambda}\, , 
\end{align}
and the minimum value is 
\begin{align}
\label{diffPSDaddsqueeze}
&\mathcal{N}[\nu,\mu_\lambda,\lambda] = \\
&\E^{-\lambda}\sqrt{\frac{3}{2}\left(1+\frac{3}{2}\sinh^2 \! \lambda\right)}+\E^{-\lambda}\cosh\lambda\,\text{sgn}(\nu) \,. \nonumber
\end{align}

For $\nu>0$, 
$\mathcal{N}[\nu,\mu_\lambda,\lambda]$ \eqref{diffPSDaddsqueeze} is decreasing in~$\lambda$, it is below the SQL value $\sqrt{3/2}+1$ \eqref{diffPSDaddSQL} for $\lambda>0$, and as $\lambda\to\infty$ it approaches~$5/4$. For $\nu<0$, 
$\mathcal{N}[\nu,\mu_\lambda,\lambda]$
is decreasing in $\lambda$ for $\lambda < \lambda_0$ and increasing in $\lambda$ for $\lambda > \lambda_0$, attaining its minimum at $\lambda_0 = \frac12\ln\bigl(5+4\sqrt{10}\bigr)-\ln3 \approx 0.3367$, with the minimum value $\frac16\bigl(\sqrt{10}-2\bigr)  \approx 0.1937$, 
and approaching $1/4$ as $\lambda\to\infty$. 
$\mathcal{N}[\nu,\mu_\lambda,\lambda]$ is below the SQL value $\sqrt{3/2}-1$ \eqref{diffPSDaddSQL} for $0<\lambda< \lambda_1=\frac12\ln\bigl(\frac{5}{23}(19+8\sqrt{6})\bigr) \approx 1.0635$. 

\begin{figure}
    \centering
    \includegraphics[width=\columnwidth]{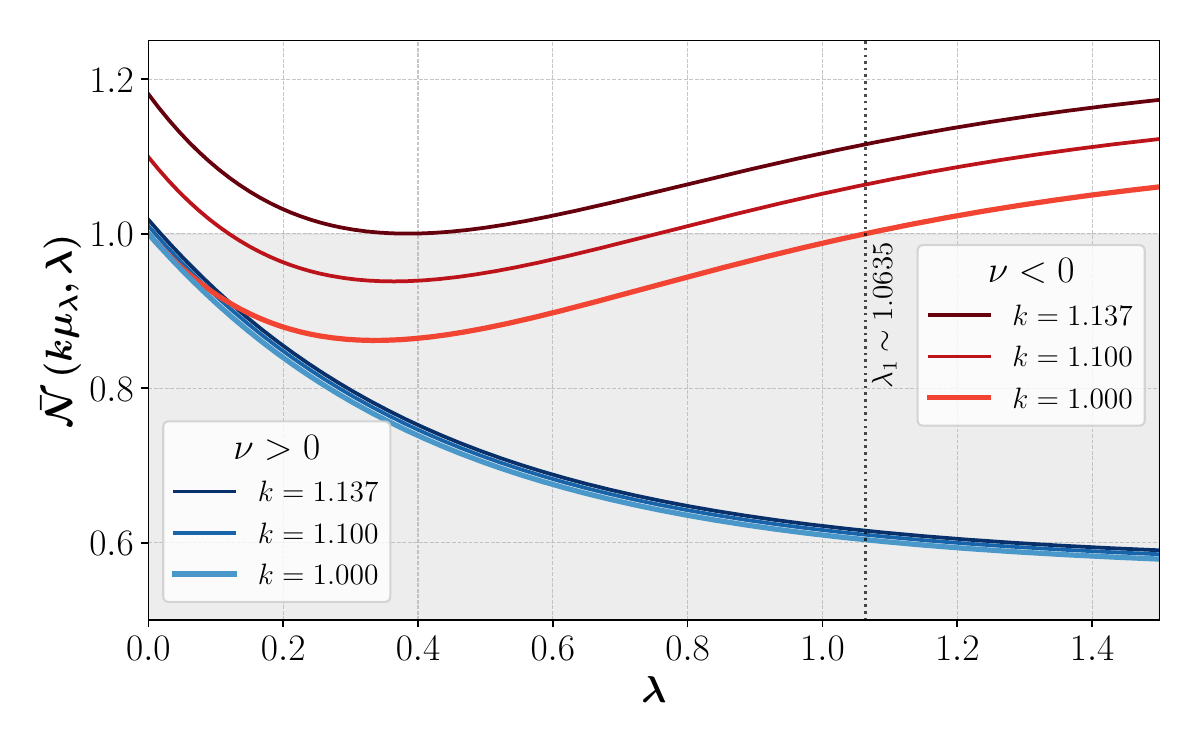}
    \caption{Plots of the normalised added noise~$\bar{\mathcal{N}}$, defined as \eqref{diffPSDadd} divided by the SQL value~\eqref{diffPSDaddSQL}, as a function of $\lambda$ (squeezing parameter) and~$\mu$ (laser-BEC coupling parameter), for $0 \le \lambda \le 1.5$ and $\mu = k\mu_\lambda$, where $\mu_\lambda$ \eqref{mulambda} is the optimal value of~$\mu$. Three selected values of $k$ are shown for both $\nu>0$ and $\nu<0$. The $k=1$ curves are~\eqref{diffPSDaddsqueeze}, 
    and the ranges of $\lambda$ where $\bar{\mathcal{N}} < 1$ are as described in the text. For $k>1$, the ranges of $\lambda$ where $\bar{\mathcal{N}} < 1$ are narrower: 
    this range exists for $1 < k < 3^{-1/2} \big(2 + 2\sqrt{6} + \sqrt{19 + 8\sqrt{6}}\,\big)^{1/2} \approx 2.091$ when $\nu>0$, 
    and for 
    $1 < k < 3^{-1/4} \big(\sqrt{6} -2 + \sqrt{13 - 4\sqrt{6}}\,\big)^{1/2} \approx 1.137$ when $\nu<0$.  
    For $0<k<1$, replace $k$ by~$1/k$. The physical mechanism that makes the added noise more sensitive to $\mu$ for $\nu<0$ than for $\nu>0$ is the same as in the backaction cooling in the measurement of zero point mechanical oscillator in a cavity optomechanical system~\cite{Khalili2012}, as we show in Supplemental Material~\cite{suppmat}.}
    \label{fig:SQL}
\end{figure}

Collecting, the added noise $\mathcal{N}[\nu,\mu_\lambda,\lambda]$ \eqref{diffPSDaddsqueeze}
can be brought below the SQL value \eqref{diffPSDaddSQL} for both $\nu>0$ and $\nu<0$ by choosing the squeezing parameter to be in the interval $0<\lambda< \lambda_1 \approx 1.0635$, when the dimensionless coupling parameter $\mu$ has the optimal value $\mu_\lambda$~\eqref{mulambda}. 
When $\mu$ deviates from the optimal value, the interval of $\lambda$ where this happens becomes narrower, in a manner shown in Figure~\ref{fig:SQL}. The squeezing range ($0<\lambda<\lambda_1$) required for simultaneous sub-SQL performance for $\nu>0$ and $\nu<0$ corresponds to $6$-$9$ dB of two-mode squeezing, well within the capabilities of fibre- and OPO-based sources~\cite{Eberle2013}.

We exemplify the achievable sensitivity of our detection scheme when operating near the SQL by estimating the threshold for resolving relative density fluctuations $\delta\rho/\rho_0$. This threshold is defined by a unit signal-to-noise ratio (SNR), $S_{\phi\phi}[\nu]/\mathcal{N}[\nu,\mu_\lambda,\lambda]=1$. In Appendix~A, we find that the field $\phi$ is related to BEC density fluctuations by $\phi=\delta\rho/(2\sqrt{m\rho_0})$. Using SI units and estimating $S_{\phi\phi}\approx\phi^2\Delta t$, we find
\begin{equation}\label{eq:sens}
    \frac{\delta\rho}{\rho_0}\approx\sqrt{\frac{4mS_{\phi\phi}[\nu]}{\hbar \rho_0\Delta t}}~=~\sqrt{\frac{4m\mathcal{N}[\nu,\mu_\lambda,\lambda]}{\hbar \rho_0\Delta t}}\,.
\end{equation}Measuring $^{133}$Cs BEC atoms for $10\mathrm{ms}$ with $\rho_0=10\mu\text{m}^{-2}$ and $\nu<0$ yields $\delta\rho/\rho_0\approx0.137$ when $\lambda=1$. Remaining at the point of minimal added noise \eqref{diffPSDaddsqueeze}, the sensitivity of this scheme can be further improved either with squeezing, repeated experimental realisations, or potentially cavity enhancement~\cite{Lueghamer2025}.

\textbf{Accelerated interaction paths---}As discussed in the text surrounding~\eqref{eqn:: a transformation} and~\eqref{eq:firstorder}, the effective relativistic field $\phi(t,\bm{x})$, corresponding to density fluctuations in the homogeneous BEC, evolves on an \textit{analogue\/} spacetime geometry determined by the bulk condensate $\Phi_0$~\cite{PhysRevLett.125.213603}, where the speed of sound in the BEC plays the role of the speed of light in a relativistic setting. It was further pointed out in~\cite{PhysRevLett.125.213603} that a localised laser---as described above---measures these fluctuations in a manner consistent with the particle-detector model known in the literature as the Unruh-DeWitt detector.

With this in mind, we apply our detection scheme to stationary accelerated interaction paths, simulating the Unruh effect---the celebrated prediction in Minkowski spacetime quantum field theory that an accelerated observer and an inertial observer measure different quantum fluctuations~\cite{Fulling,Davies1975,unruh76,Fulling:2014}. Recall that from~\eqref{eqn:: PSD} onwards, we have assumed the interaction trajectory to be stationary~\cite{LetawStationary,Bunney:2023mkh}. While the best-known version of the Unruh effect is for uniform linear acceleration \cite{Fulling,Davies1975,unruh76}, 
versions of the effect exist for all types of uniform accelerations~\cite{Letaw:1980ik,Good:2020hav}, including uniform circular motion, which has special experimental interest in that it stays in a finite spatial volume arbitrarily long. 
The Unruh effect exists also in analogue spacetimes: 
the uniform circular motion case has been recently discussed in BECs \cite{PhysRevD.102.085006,PhysRevLett.125.213603} and in superfluid helium~\cite{Bunney:2023vyj,https://doi.org/10.48550/arxiv.2302.12023}. 
We shall now specialise to a circular path, and address the information that $S_{\phi \phi}[\nu]$ carries about the circular motion Unruh effect.

The quantum vacuum fluctuations as detected by linearly accelerating observers are thermally distributed, described by a temperature proportional to the acceleration. A subtlety with the circular motion Unruh effect is that it cannot be described by a single temperature parameter. It can, however, be associated with an effective, frequency-dependent temperature $T_{\text{eff}}(\nu)$ by writing
\begin{equation}\label{eq:Teff}
\E^{\nu/T_{\text{eff}}(\nu)} ~\coloneqq~ \frac{S_{\phi\phi}[-\nu]}{S_{\phi\phi}[\nu]  }\,.
\end{equation}
This amounts to matching the ratio of the excitation and deexcitation rates~\eqref{eqn:: PSD} along the accelerated path to Einstein's detailed balance formula. For circular motion, $T_{\text{eff}}$ is approximately constant and of the same order of magnitude as the predicted linear-acceleration Unruh temperature over most of the accessible parameter space~\cite{Unruh:1998gq,PhysRevD.102.085006}.

In the experimental proposal \cite{PhysRevLett.125.213603} for realising the circular motion Unruh effect, the physical system was a BEC comprised of $^{133}$Cs atoms with a two-dimensional number density $\rho_0=10^3\mu\mrm{m}^{-2}$, and a laser of frequency $\omega_0/(2\pi)=10^{14}\mathrm{Hz}$ with a beam width $r_0=3\mu\mathrm{m}$. 
Upon fixing the BEC parameters, the chosen squeezing $\lambda$ defines the power and detuning relations of the laser. Working at the point of minimal backaction, the sensitivity within the phononic theory can only be improved by increasing the statistics or the total measurement time (analysis bandwidth)~\cite{PhysRevLett.125.213603}. The non-destructiveness of the measurement beyond the phononic theory has to be ensured \emph{a posteriori}; first a sufficiently long lifetime of the BEC to resolve phononic frequencies, given typical experimental heating and atom-loss. Second a sufficiently small incoherent scattering rate $\Gamma_{sc}=4\hat{\alpha}_I\bar{P}/(\pi r_0^2)$ connected to the imaginary part of the atomic polarisability $\hat{\alpha}_I$, which is beyond the presented phononic theory but completely defined by it. Here, $\bar{P}\approx 2\omega_0\alpha^2$ is the laser power entering the BEC, averaged over modulation cycles.
Working at the SQL with $\alpha=\alpha_{\mathrm{SQL}}$, 
which corresponds to $\mu^2=2\sqrt{2/3}$, we find a photon scattering rate $\Gamma_{\mathrm{sc}}\approx 0.0020\text{Hz}$, which is much smaller than unity. In the case of an initially squeezed state, we show in Supplemental Material \cite{suppmat} that $\Gamma_{\mathrm{sc}}\approx0.0031\text{Hz}$ when $\mu$ takes the optimal value \eqref{mulambda} and $\lambda$ is set to the minimum of the red curve in Figure~\ref{fig:SQL}. As both scattering rates are considerably smaller than unity (nondestructive) and the typical frequencies of the phonons of interest, we conclude that the detection scheme presented in this Letter represents a promising and feasible approach for an experimental realisation of the circular motion Unruh effect:
the probe beam 
acts as an Unruh-DeWitt detector \textit{beyond the standard quantum limit}.

\textbf{Discussion---}Traditionally, optomechanics deals with single- or few-mode mechanical systems, such as membranes or mirrors, and this approach has been extended to BECs and superfluids coupled to optical cavities~\cite{PhysRevLett.108.133601,PhysRevLett.122.153601,Gupta:2024mhs}. In these systems, the optomechanical interaction typically involves a single collective excitation, such as the centre-of-mass or breathing mode, which acts as the effective mechanical degree of freedom. By contrast, our approach expands the optomechanical toolbox to a quantum fluid probed with laser beams along arbitrary stationary spatial paths, enabling access to unequal-time density correlations in a multi-mode BEC environment. 

By employing a two-tone heterodyne optical readout and systematically analysing both imprecision and backaction noise, we provide a theoretical framework for extracting unequal-time correlation functions with sensitivity reaching, and potentially surpassing, the shot-noise limit of the probe field. This opens up investigations of fundamental properties of the smallest excitations of a quantum fluid, including the quantum vacuum state of the system and its observer-dependence. 
Our detection scheme finds its optimal sensitivity by balancing the quantum imprints left by the probe beam against the disturbance it causes, allowing the wave dynamics to be recorded with maximal fidelity. 

While our concrete implementation focuses on a~BEC, with the experimentally realistic parameters provided in the text below \eqref{eq:sens} and in the last paragraph of the Accelerated interaction paths section, the scheme is adaptable to other quantum fluids, placing nondestructive access to unequal-time correlations in quantum fluids in general reach.

\textbf{Acknowledgements---}We thank Chris Goodwin for generating Figure~\ref{fig:exp}. Partial support for the work of CG was provided by the European Research Council under the Consolidator Grant COQCOoN (Grant No.\ 820079). CRDB gratefully acknowledges the support of the Engineering and Physical Sciences Research Council (EP/W524402/1). JS acknowledges the support provided by the European Research Council, ERC-AdG \textit{Emergence in Quantum Physics} (EmQ) under Grant Agreement No.\ 101097858. The work of JL was supported by United Kingdom Research and Innovation Science and Technology Facilities Council [grant numbers ST/S002227/1, ST/T006900/1 and ST/Y004523/1], and has benefited from the activities of COST Action CA23115:
Relativistic Quantum Information, funded  by COST (European Cooperation
in Science and Technology). SW acknowledges the support provided
by the Leverhulme Research Leadership Award (RL2019-020), the Royal Society University Research Fellowship (UF120112, RF/ERE/210198, RGF/EA/180286, 
RGF/EA/181015), and partial support by the Science
and Technology Facilities Council (Theory Consolidated Grant ST/P000703/1), the Science and Technology Facilities Council on Quantum Simulators for Fundamental Physics (ST/T006900/1) as part of the UKRI Quantum Technologies for Fundamental Physics programme. WU thanks the
Natural Sciences and Engineering Research Council of
Canada (NSERC) (Grant No.\ 5-80441). For the purpose of open access, the authors have applied a CC BY public copyright licence to any Author Accepted Manuscript version arising.

\bibliography{bibliography}

\begin{thebibliography}{60}%
\makeatletter
\providecommand \@ifxundefined [1]{%
 \@ifx{#1\undefined}
}%
\providecommand \@ifnum [1]{%
 \ifnum #1\expandafter \@firstoftwo
 \else \expandafter \@secondoftwo
 \fi
}%
\providecommand \@ifx [1]{%
 \ifx #1\expandafter \@firstoftwo
 \else \expandafter \@secondoftwo
 \fi
}%
\providecommand \natexlab [1]{#1}%
\providecommand \enquote  [1]{``#1''}%
\providecommand \bibnamefont  [1]{#1}%
\providecommand \bibfnamefont [1]{#1}%
\providecommand \citenamefont [1]{#1}%
\providecommand \href@noop [0]{\@secondoftwo}%
\providecommand \href [0]{\begingroup \@sanitize@url \@href}%
\providecommand \@href[1]{\@@startlink{#1}\@@href}%
\providecommand \@@href[1]{\endgroup#1\@@endlink}%
\providecommand \@sanitize@url [0]{\catcode `\\12\catcode `\$12\catcode
  `\&12\catcode `\#12\catcode `\^12\catcode `\_12\catcode `\%12\relax}%
\providecommand \@@startlink[1]{}%
\providecommand \@@endlink[0]{}%
\providecommand \url  [0]{\begingroup\@sanitize@url \@url }%
\providecommand \@url [1]{\endgroup\@href {#1}{\urlprefix }}%
\providecommand \urlprefix  [0]{URL }%
\providecommand \Eprint [0]{\href }%
\providecommand \doibase [0]{https://doi.org/}%
\providecommand \selectlanguage [0]{\@gobble}%
\providecommand \bibinfo  [0]{\@secondoftwo}%
\providecommand \bibfield  [0]{\@secondoftwo}%
\providecommand \translation [1]{[#1]}%
\providecommand \BibitemOpen [0]{}%
\providecommand \bibitemStop [0]{}%
\providecommand \bibitemNoStop [0]{.\EOS\space}%
\providecommand \EOS [0]{\spacefactor3000\relax}%
\providecommand \BibitemShut  [1]{\csname bibitem#1\endcsname}%
\let\auto@bib@innerbib\@empty
\bibitem [{\citenamefont {Tajik}\ \emph
  {et~al.}(2023{\natexlab{a}})\citenamefont {Tajik}, \citenamefont {Kukuljan},
  \citenamefont {Sotiriadis}, \citenamefont {Rauer}, \citenamefont
  {Schweigler}, \citenamefont {Cataldini}, \citenamefont {Sabino},
  \citenamefont {M{\o}ller}, \citenamefont {Sch{\"u}ttelkopf}, \citenamefont
  {Ji}, \citenamefont {Sels}, \citenamefont {Demler},\ and\ \citenamefont
  {Schmiedmayer}}]{Tajik2023_AreaLaw}%
  \BibitemOpen
  \bibfield  {author} {\bibinfo {author} {\bibfnamefont {M.}~\bibnamefont
  {Tajik}}, \bibinfo {author} {\bibfnamefont {I.}~\bibnamefont {Kukuljan}},
  \bibinfo {author} {\bibfnamefont {S.}~\bibnamefont {Sotiriadis}}, \bibinfo
  {author} {\bibfnamefont {B.}~\bibnamefont {Rauer}}, \bibinfo {author}
  {\bibfnamefont {T.}~\bibnamefont {Schweigler}}, \bibinfo {author}
  {\bibfnamefont {F.}~\bibnamefont {Cataldini}}, \bibinfo {author}
  {\bibfnamefont {J.}~\bibnamefont {Sabino}}, \bibinfo {author} {\bibfnamefont
  {F.}~\bibnamefont {M{\o}ller}}, \bibinfo {author} {\bibfnamefont
  {P.}~\bibnamefont {Sch{\"u}ttelkopf}}, \bibinfo {author} {\bibfnamefont
  {S.}~\bibnamefont {Ji}}, \bibinfo {author} {\bibfnamefont {D.}~\bibnamefont
  {Sels}}, \bibinfo {author} {\bibfnamefont {E.}~\bibnamefont {Demler}},\ and\
  \bibinfo {author} {\bibfnamefont {J.}~\bibnamefont {Schmiedmayer}},\
  }\bibfield  {title} {\bibinfo {title} {Verification of the area law of mutual
  information in a quantum field simulator},\ }\href
  {https://doi.org/10.1038/s41567-023-02027-1} {\bibfield  {journal} {\bibinfo
  {journal} {Nature Physics}\ }\textbf {\bibinfo {volume} {19}},\ \bibinfo
  {pages} {1022} (\bibinfo {year} {2023}{\natexlab{a}})},\ \Eprint
  {https://arxiv.org/abs/2206.10563} {arXiv:2206.10563 [cond-mat.quant-gas]}
  \BibitemShut {NoStop}%
\bibitem [{\citenamefont {Dogra}\ \emph {et~al.}(2023)\citenamefont {Dogra},
  \citenamefont {Martirosyan}, \citenamefont {Hilker}, \citenamefont {Glidden},
  \citenamefont {Etrych}, \citenamefont {Cao}, \citenamefont {Eigen},
  \citenamefont {Smith},\ and\ \citenamefont {Hadzibabic}}]{Dogra2023}%
  \BibitemOpen
  \bibfield  {author} {\bibinfo {author} {\bibfnamefont {L.~H.}\ \bibnamefont
  {Dogra}}, \bibinfo {author} {\bibfnamefont {G.}~\bibnamefont {Martirosyan}},
  \bibinfo {author} {\bibfnamefont {T.~A.}\ \bibnamefont {Hilker}}, \bibinfo
  {author} {\bibfnamefont {J.~A.~P.}\ \bibnamefont {Glidden}}, \bibinfo
  {author} {\bibfnamefont {J.}~\bibnamefont {Etrych}}, \bibinfo {author}
  {\bibfnamefont {A.}~\bibnamefont {Cao}}, \bibinfo {author} {\bibfnamefont
  {C.}~\bibnamefont {Eigen}}, \bibinfo {author} {\bibfnamefont {R.~P.}\
  \bibnamefont {Smith}},\ and\ \bibinfo {author} {\bibfnamefont
  {Z.}~\bibnamefont {Hadzibabic}},\ }\bibfield  {title} {\bibinfo {title}
  {Universal equation of state for wave turbulence in a quantum gas},\ }\href
  {https://doi.org/10.1038/s41586-023-06240-z} {\bibfield  {journal} {\bibinfo
  {journal} {Nature}\ }\textbf {\bibinfo {volume} {620}},\ \bibinfo {pages}
  {521} (\bibinfo {year} {2023})}\BibitemShut {NoStop}%
\bibitem [{\citenamefont {White}\ \emph {et~al.}(2010)\citenamefont {White},
  \citenamefont {Barenghi}, \citenamefont {Proukakis}, \citenamefont {Youd},\
  and\ \citenamefont {Wacks}}]{PhysRevLett.104.075301}%
  \BibitemOpen
  \bibfield  {author} {\bibinfo {author} {\bibfnamefont {A.~C.}\ \bibnamefont
  {White}}, \bibinfo {author} {\bibfnamefont {C.~F.}\ \bibnamefont {Barenghi}},
  \bibinfo {author} {\bibfnamefont {N.~P.}\ \bibnamefont {Proukakis}}, \bibinfo
  {author} {\bibfnamefont {A.~J.}\ \bibnamefont {Youd}},\ and\ \bibinfo
  {author} {\bibfnamefont {D.~H.}\ \bibnamefont {Wacks}},\ }\bibfield  {title}
  {\bibinfo {title} {Nonclassical velocity statistics in a turbulent atomic
  {Bose-Einstein} condensate},\ }\href
  {https://doi.org/10.1103/PhysRevLett.104.075301} {\bibfield  {journal}
  {\bibinfo  {journal} {Phys. Rev. Lett.}\ }\textbf {\bibinfo {volume} {104}},\
  \bibinfo {pages} {075301} (\bibinfo {year} {2010})}\BibitemShut {NoStop}%
\bibitem [{\citenamefont {Henn}\ \emph {et~al.}(2009)\citenamefont {Henn},
  \citenamefont {Seman}, \citenamefont {Roati}, \citenamefont {Magalh\~aes},\
  and\ \citenamefont {Bagnato}}]{PhysRevLett.103.045301}%
  \BibitemOpen
  \bibfield  {author} {\bibinfo {author} {\bibfnamefont {E.~A.~L.}\
  \bibnamefont {Henn}}, \bibinfo {author} {\bibfnamefont {J.~A.}\ \bibnamefont
  {Seman}}, \bibinfo {author} {\bibfnamefont {G.}~\bibnamefont {Roati}},
  \bibinfo {author} {\bibfnamefont {K.~M.~F.}\ \bibnamefont {Magalh\~aes}},\
  and\ \bibinfo {author} {\bibfnamefont {V.~S.}\ \bibnamefont {Bagnato}},\
  }\bibfield  {title} {\bibinfo {title} {Emergence of turbulence in an
  oscillating {Bose-Einstein} condensate},\ }\href
  {https://doi.org/10.1103/PhysRevLett.103.045301} {\bibfield  {journal}
  {\bibinfo  {journal} {Phys. Rev. Lett.}\ }\textbf {\bibinfo {volume} {103}},\
  \bibinfo {pages} {045301} (\bibinfo {year} {2009})}\BibitemShut {NoStop}%
\bibitem [{\citenamefont {Parker}\ and\ \citenamefont
  {Adams}(2005)}]{PhysRevLett.95.145301}%
  \BibitemOpen
  \bibfield  {author} {\bibinfo {author} {\bibfnamefont {N.~G.}\ \bibnamefont
  {Parker}}\ and\ \bibinfo {author} {\bibfnamefont {C.~S.}\ \bibnamefont
  {Adams}},\ }\bibfield  {title} {\bibinfo {title} {Emergence and decay of
  turbulence in stirred atomic {Bose-Einstein} condensates},\ }\href
  {https://doi.org/10.1103/PhysRevLett.95.145301} {\bibfield  {journal}
  {\bibinfo  {journal} {Phys. Rev. Lett.}\ }\textbf {\bibinfo {volume} {95}},\
  \bibinfo {pages} {145301} (\bibinfo {year} {2005})}\BibitemShut {NoStop}%
\bibitem [{\citenamefont {Opanchuk}\ \emph {et~al.}(2013)\citenamefont
  {Opanchuk}, \citenamefont {Polkinghorne}, \citenamefont {Fialko},
  \citenamefont {Brand},\ and\ \citenamefont {Drummond}}]{Opanchuk2013}%
  \BibitemOpen
  \bibfield  {author} {\bibinfo {author} {\bibfnamefont {B.}~\bibnamefont
  {Opanchuk}}, \bibinfo {author} {\bibfnamefont {R.}~\bibnamefont
  {Polkinghorne}}, \bibinfo {author} {\bibfnamefont {O.}~\bibnamefont
  {Fialko}}, \bibinfo {author} {\bibfnamefont {J.}~\bibnamefont {Brand}},\ and\
  \bibinfo {author} {\bibfnamefont {P.~D.}\ \bibnamefont {Drummond}},\
  }\bibfield  {title} {\bibinfo {title} {{Quantum simulations of the early
  universe}},\ }\href {https://doi.org/10.1002/andp.201300113} {\bibfield
  {journal} {\bibinfo  {journal} {Annalen Phys.}\ }\textbf {\bibinfo {volume}
  {525}},\ \bibinfo {pages} {866} (\bibinfo {year} {2013})},\ \Eprint
  {https://arxiv.org/abs/1305.5314} {arXiv:1305.5314 [cond-mat.quant-gas]}
  \BibitemShut {NoStop}%
\bibitem [{\citenamefont {Fialko}\ \emph {et~al.}(2015)\citenamefont {Fialko},
  \citenamefont {Opanchuk}, \citenamefont {Sidorov}, \citenamefont {Drummond},\
  and\ \citenamefont {Brand}}]{Fialko2015}%
  \BibitemOpen
  \bibfield  {author} {\bibinfo {author} {\bibfnamefont {O.}~\bibnamefont
  {Fialko}}, \bibinfo {author} {\bibfnamefont {B.}~\bibnamefont {Opanchuk}},
  \bibinfo {author} {\bibfnamefont {A.~I.}\ \bibnamefont {Sidorov}}, \bibinfo
  {author} {\bibfnamefont {P.~D.}\ \bibnamefont {Drummond}},\ and\ \bibinfo
  {author} {\bibfnamefont {J.}~\bibnamefont {Brand}},\ }\bibfield  {title}
  {\bibinfo {title} {{Fate of the false vacuum: towards realization with
  ultra-cold atoms}},\ }\href {https://doi.org/10.1209/0295-5075/110/56001}
  {\bibfield  {journal} {\bibinfo  {journal} {EPL}\ }\textbf {\bibinfo {volume}
  {110}},\ \bibinfo {pages} {56001} (\bibinfo {year} {2015})},\ \Eprint
  {https://arxiv.org/abs/1408.1163} {arXiv:1408.1163 [cond-mat.quant-gas]}
  \BibitemShut {NoStop}%
\bibitem [{\citenamefont {Fialko}\ \emph {et~al.}(2017)\citenamefont {Fialko},
  \citenamefont {Opanchuk}, \citenamefont {Sidorov}, \citenamefont {Drummond},\
  and\ \citenamefont {Brand}}]{Fialko2017}%
  \BibitemOpen
  \bibfield  {author} {\bibinfo {author} {\bibfnamefont {O.}~\bibnamefont
  {Fialko}}, \bibinfo {author} {\bibfnamefont {B.}~\bibnamefont {Opanchuk}},
  \bibinfo {author} {\bibfnamefont {A.~I.}\ \bibnamefont {Sidorov}}, \bibinfo
  {author} {\bibfnamefont {P.~D.}\ \bibnamefont {Drummond}},\ and\ \bibinfo
  {author} {\bibfnamefont {J.}~\bibnamefont {Brand}},\ }\bibfield  {title}
  {\bibinfo {title} {{The universe on a table top: engineering quantum decay of
  a relativistic scalar field from a metastable vacuum}},\ }\href
  {https://doi.org/10.1088/1361-6455/50/2/024003} {\bibfield  {journal}
  {\bibinfo  {journal} {J. Phys. B}\ }\textbf {\bibinfo {volume} {50}},\
  \bibinfo {pages} {024003} (\bibinfo {year} {2017})},\ \Eprint
  {https://arxiv.org/abs/1607.01460} {arXiv:1607.01460 [cond-mat.quant-gas]}
  \BibitemShut {NoStop}%
\bibitem [{\citenamefont {Braden}\ \emph {et~al.}(2018)\citenamefont {Braden},
  \citenamefont {Johnson}, \citenamefont {Peiris},\ and\ \citenamefont
  {Weinfurtner}}]{Braden2018}%
  \BibitemOpen
  \bibfield  {author} {\bibinfo {author} {\bibfnamefont {J.}~\bibnamefont
  {Braden}}, \bibinfo {author} {\bibfnamefont {M.~C.}\ \bibnamefont {Johnson}},
  \bibinfo {author} {\bibfnamefont {H.~V.}\ \bibnamefont {Peiris}},\ and\
  \bibinfo {author} {\bibfnamefont {S.}~\bibnamefont {Weinfurtner}},\
  }\bibfield  {title} {\bibinfo {title} {{Towards the cold atom analog false
  vacuum}},\ }\href {https://doi.org/10.1007/JHEP07(2018)014} {\bibfield
  {journal} {\bibinfo  {journal} {JHEP}\ }\textbf {\bibinfo {volume} {07}},\
  \bibinfo {pages} {014}},\ \Eprint {https://arxiv.org/abs/1712.02356}
  {arXiv:1712.02356 [hep-th]} \BibitemShut {NoStop}%
\bibitem [{\citenamefont {Billam}\ \emph {et~al.}(2019)\citenamefont {Billam},
  \citenamefont {Gregory}, \citenamefont {Michel},\ and\ \citenamefont
  {Moss}}]{Billam2019PRL}%
  \BibitemOpen
  \bibfield  {author} {\bibinfo {author} {\bibfnamefont {T.~P.}\ \bibnamefont
  {Billam}}, \bibinfo {author} {\bibfnamefont {R.}~\bibnamefont {Gregory}},
  \bibinfo {author} {\bibfnamefont {F.}~\bibnamefont {Michel}},\ and\ \bibinfo
  {author} {\bibfnamefont {I.~G.}\ \bibnamefont {Moss}},\ }\bibfield  {title}
  {\bibinfo {title} {{Simulating seeded vacuum decay in a cold atom system}},\
  }\href {https://doi.org/10.1103/PhysRevD.100.065016} {\bibfield  {journal}
  {\bibinfo  {journal} {Phys. Rev. D}\ }\textbf {\bibinfo {volume} {100}},\
  \bibinfo {pages} {065016} (\bibinfo {year} {2019})},\ \Eprint
  {https://arxiv.org/abs/1811.09169} {arXiv:1811.09169 [hep-th]} \BibitemShut
  {NoStop}%
\bibitem [{\citenamefont {Braden}\ \emph {et~al.}(2019)\citenamefont {Braden},
  \citenamefont {Johnson}, \citenamefont {Peiris}, \citenamefont {Pontzen},\
  and\ \citenamefont {Weinfurtner}}]{Braden2019}%
  \BibitemOpen
  \bibfield  {author} {\bibinfo {author} {\bibfnamefont {J.}~\bibnamefont
  {Braden}}, \bibinfo {author} {\bibfnamefont {M.~C.}\ \bibnamefont {Johnson}},
  \bibinfo {author} {\bibfnamefont {H.~V.}\ \bibnamefont {Peiris}}, \bibinfo
  {author} {\bibfnamefont {A.}~\bibnamefont {Pontzen}},\ and\ \bibinfo {author}
  {\bibfnamefont {S.}~\bibnamefont {Weinfurtner}},\ }\bibfield  {title}
  {\bibinfo {title} {{Nonlinear Dynamics of the Cold Atom Analog False
  Vacuum}},\ }\href {https://doi.org/10.1007/JHEP10(2019)174} {\bibfield
  {journal} {\bibinfo  {journal} {JHEP}\ }\textbf {\bibinfo {volume} {10}},\
  \bibinfo {pages} {174}},\ \Eprint {https://arxiv.org/abs/1904.07873}
  {arXiv:1904.07873 [hep-th]} \BibitemShut {NoStop}%
\bibitem [{\citenamefont {Ng}\ \emph {et~al.}(2021)\citenamefont {Ng},
  \citenamefont {Opanchuk}, \citenamefont {Thenabadu}, \citenamefont {Reid},\
  and\ \citenamefont {Drummond}}]{Ng2021}%
  \BibitemOpen
  \bibfield  {author} {\bibinfo {author} {\bibfnamefont {K.~L.}\ \bibnamefont
  {Ng}}, \bibinfo {author} {\bibfnamefont {B.}~\bibnamefont {Opanchuk}},
  \bibinfo {author} {\bibfnamefont {M.}~\bibnamefont {Thenabadu}}, \bibinfo
  {author} {\bibfnamefont {M.}~\bibnamefont {Reid}},\ and\ \bibinfo {author}
  {\bibfnamefont {P.~D.}\ \bibnamefont {Drummond}},\ }\bibfield  {title}
  {\bibinfo {title} {{Fate of the False Vacuum: Finite Temperature, Entropy,
  and Topological Phase in Quantum Simulations of the Early Universe}},\ }\href
  {https://doi.org/10.1103/PRXQuantum.2.010350} {\bibfield  {journal} {\bibinfo
   {journal} {PRX Quantum}\ }\textbf {\bibinfo {volume} {2}},\ \bibinfo {pages}
  {010350} (\bibinfo {year} {2021})},\ \Eprint
  {https://arxiv.org/abs/2010.08665} {arXiv:2010.08665 [quant-ph]} \BibitemShut
  {NoStop}%
\bibitem [{\citenamefont {Jenkins}\ \emph
  {et~al.}(2024{\natexlab{a}})\citenamefont {Jenkins}, \citenamefont {Braden},
  \citenamefont {Peiris}, \citenamefont {Pontzen}, \citenamefont {Johnson},\
  and\ \citenamefont {Weinfurtner}}]{Jenkins2024a}%
  \BibitemOpen
  \bibfield  {author} {\bibinfo {author} {\bibfnamefont {A.~C.}\ \bibnamefont
  {Jenkins}}, \bibinfo {author} {\bibfnamefont {J.}~\bibnamefont {Braden}},
  \bibinfo {author} {\bibfnamefont {H.~V.}\ \bibnamefont {Peiris}}, \bibinfo
  {author} {\bibfnamefont {A.}~\bibnamefont {Pontzen}}, \bibinfo {author}
  {\bibfnamefont {M.~C.}\ \bibnamefont {Johnson}},\ and\ \bibinfo {author}
  {\bibfnamefont {S.}~\bibnamefont {Weinfurtner}},\ }\bibfield  {title}
  {\bibinfo {title} {{Analog vacuum decay from vacuum initial conditions}},\
  }\href {https://doi.org/10.1103/PhysRevD.109.023506} {\bibfield  {journal}
  {\bibinfo  {journal} {Phys. Rev. D}\ }\textbf {\bibinfo {volume} {109}},\
  \bibinfo {pages} {023506} (\bibinfo {year} {2024}{\natexlab{a}})},\ \Eprint
  {https://arxiv.org/abs/2307.02549} {arXiv:2307.02549 [cond-mat.quant-gas]}
  \BibitemShut {NoStop}%
\bibitem [{\citenamefont {Zenesini}\ \emph {et~al.}(2024)\citenamefont
  {Zenesini}, \citenamefont {Berti}, \citenamefont {Cominotti}, \citenamefont
  {Rogora}, \citenamefont {Moss}, \citenamefont {Billam}, \citenamefont
  {Carusotto}, \citenamefont {Lamporesi}, \citenamefont {Recati},\ and\
  \citenamefont {Ferrari}}]{Zenesini2024}%
  \BibitemOpen
  \bibfield  {author} {\bibinfo {author} {\bibfnamefont {A.}~\bibnamefont
  {Zenesini}}, \bibinfo {author} {\bibfnamefont {A.}~\bibnamefont {Berti}},
  \bibinfo {author} {\bibfnamefont {R.}~\bibnamefont {Cominotti}}, \bibinfo
  {author} {\bibfnamefont {C.}~\bibnamefont {Rogora}}, \bibinfo {author}
  {\bibfnamefont {I.~G.}\ \bibnamefont {Moss}}, \bibinfo {author}
  {\bibfnamefont {T.~P.}\ \bibnamefont {Billam}}, \bibinfo {author}
  {\bibfnamefont {I.}~\bibnamefont {Carusotto}}, \bibinfo {author}
  {\bibfnamefont {G.}~\bibnamefont {Lamporesi}}, \bibinfo {author}
  {\bibfnamefont {A.}~\bibnamefont {Recati}},\ and\ \bibinfo {author}
  {\bibfnamefont {G.}~\bibnamefont {Ferrari}},\ }\bibfield  {title} {\bibinfo
  {title} {{False vacuum decay via bubble formation in ferromagnetic
  superfluids}},\ }\href {https://doi.org/10.1038/s41567-023-02345-4}
  {\bibfield  {journal} {\bibinfo  {journal} {Nature Phys.}\ }\textbf {\bibinfo
  {volume} {20}},\ \bibinfo {pages} {558} (\bibinfo {year} {2024})},\ \Eprint
  {https://arxiv.org/abs/2305.05225} {arXiv:2305.05225 [hep-ph]} \BibitemShut
  {NoStop}%
\bibitem [{\citenamefont {Cominotti}\ \emph {et~al.}(2025)\citenamefont
  {Cominotti}, \citenamefont {Baroni}, \citenamefont {Rogora}, \citenamefont
  {Andreoni}, \citenamefont {Guarda}, \citenamefont {Lamporesi}, \citenamefont
  {Ferrari},\ and\ \citenamefont {Zenesini}}]{Cominotti2025}%
  \BibitemOpen
  \bibfield  {author} {\bibinfo {author} {\bibfnamefont {R.}~\bibnamefont
  {Cominotti}}, \bibinfo {author} {\bibfnamefont {C.}~\bibnamefont {Baroni}},
  \bibinfo {author} {\bibfnamefont {C.}~\bibnamefont {Rogora}}, \bibinfo
  {author} {\bibfnamefont {D.}~\bibnamefont {Andreoni}}, \bibinfo {author}
  {\bibfnamefont {G.}~\bibnamefont {Guarda}}, \bibinfo {author} {\bibfnamefont
  {G.}~\bibnamefont {Lamporesi}}, \bibinfo {author} {\bibfnamefont
  {G.}~\bibnamefont {Ferrari}},\ and\ \bibinfo {author} {\bibfnamefont
  {A.}~\bibnamefont {Zenesini}},\ }\bibfield  {title} {\bibinfo {title}
  {Observation of temperature effects on false vacuum decay in atomic quantum
  gases},\ }\href {https://doi.org/10.1103/l396-yysb} {\bibfield  {journal}
  {\bibinfo  {journal} {Phys. Rev. Lett.}\ }\textbf {\bibinfo {volume} {135}},\
  \bibinfo {pages} {183401} (\bibinfo {year} {2025})},\ \Eprint
  {https://arxiv.org/abs/2504.03528} {arXiv:2504.03528 [cond-mat.quant-gas]}
  \BibitemShut {NoStop}%
\bibitem [{\citenamefont {Fischer}\ and\ \citenamefont
  {Sch\"utzhold}(2004)}]{UweAndRalf2004}%
  \BibitemOpen
  \bibfield  {author} {\bibinfo {author} {\bibfnamefont {U.~R.}\ \bibnamefont
  {Fischer}}\ and\ \bibinfo {author} {\bibfnamefont {R.}~\bibnamefont
  {Sch\"utzhold}},\ }\bibfield  {title} {\bibinfo {title} {Quantum simulation
  of cosmic inflation in two-component {Bose-Einstein} condensates},\ }\href
  {https://doi.org/10.1103/PhysRevA.70.063615} {\bibfield  {journal} {\bibinfo
  {journal} {Phys. Rev. A}\ }\textbf {\bibinfo {volume} {70}},\ \bibinfo
  {pages} {063615} (\bibinfo {year} {2004})},\ \Eprint
  {https://arxiv.org/abs/cond-mat/0406470} {arXiv:cond-mat/0406470
  [cond-mat.other]} \BibitemShut {NoStop}%
\bibitem [{\citenamefont {Jaskula}\ \emph {et~al.}(2012)\citenamefont
  {Jaskula}, \citenamefont {Partridge}, \citenamefont {Bonneau}, \citenamefont
  {Lopes}, \citenamefont {Ruaudel}, \citenamefont {Boiron},\ and\ \citenamefont
  {Westbrook}}]{Jaskula2012}%
  \BibitemOpen
  \bibfield  {author} {\bibinfo {author} {\bibfnamefont {J.-C.}\ \bibnamefont
  {Jaskula}}, \bibinfo {author} {\bibfnamefont {G.~B.}\ \bibnamefont
  {Partridge}}, \bibinfo {author} {\bibfnamefont {M.}~\bibnamefont {Bonneau}},
  \bibinfo {author} {\bibfnamefont {R.}~\bibnamefont {Lopes}}, \bibinfo
  {author} {\bibfnamefont {J.}~\bibnamefont {Ruaudel}}, \bibinfo {author}
  {\bibfnamefont {D.}~\bibnamefont {Boiron}},\ and\ \bibinfo {author}
  {\bibfnamefont {C.~I.}\ \bibnamefont {Westbrook}},\ }\bibfield  {title}
  {\bibinfo {title} {An acoustic analog to the dynamical casimir effect in a
  bose-einstein condensate},\ }\href
  {https://doi.org/10.1103/PhysRevLett.109.220401} {\bibfield  {journal}
  {\bibinfo  {journal} {Phys. Rev. Lett.}\ }\textbf {\bibinfo {volume} {109}},\
  \bibinfo {pages} {220401} (\bibinfo {year} {2012})},\ \Eprint
  {https://arxiv.org/abs/1207.1338} {arXiv:1207.1338 [cond-mat.quant-gas]}
  \BibitemShut {NoStop}%
\bibitem [{\citenamefont {Viermann}\ \emph {et~al.}(2022)\citenamefont
  {Viermann}, \citenamefont {Sparn}, \citenamefont {Liebster}, \citenamefont
  {Hans}, \citenamefont {Kath}, \citenamefont {Parra-L{\'o}pez}, \citenamefont
  {Tolosa-Sime{\'o}n}, \citenamefont {S{\'a}nchez-Kuntz}, \citenamefont {Haas},
  \citenamefont {Strobel}, \citenamefont {Floerchinger},\ and\ \citenamefont
  {Oberthaler}}]{Viermann2022}%
  \BibitemOpen
  \bibfield  {author} {\bibinfo {author} {\bibfnamefont {C.}~\bibnamefont
  {Viermann}}, \bibinfo {author} {\bibfnamefont {M.}~\bibnamefont {Sparn}},
  \bibinfo {author} {\bibfnamefont {N.}~\bibnamefont {Liebster}}, \bibinfo
  {author} {\bibfnamefont {M.}~\bibnamefont {Hans}}, \bibinfo {author}
  {\bibfnamefont {E.}~\bibnamefont {Kath}}, \bibinfo {author} {\bibfnamefont
  {{\'A}.}~\bibnamefont {Parra-L{\'o}pez}}, \bibinfo {author} {\bibfnamefont
  {M.}~\bibnamefont {Tolosa-Sime{\'o}n}}, \bibinfo {author} {\bibfnamefont
  {N.}~\bibnamefont {S{\'a}nchez-Kuntz}}, \bibinfo {author} {\bibfnamefont
  {T.}~\bibnamefont {Haas}}, \bibinfo {author} {\bibfnamefont {H.}~\bibnamefont
  {Strobel}}, \bibinfo {author} {\bibfnamefont {S.}~\bibnamefont
  {Floerchinger}},\ and\ \bibinfo {author} {\bibfnamefont {M.~K.}\ \bibnamefont
  {Oberthaler}},\ }\bibfield  {title} {\bibinfo {title} {Quantum field
  simulator for dynamics in curved spacetime},\ }\href
  {https://doi.org/10.1038/s41586-022-05313-9} {\bibfield  {journal} {\bibinfo
  {journal} {Nature}\ }\textbf {\bibinfo {volume} {611}},\ \bibinfo {pages}
  {260} (\bibinfo {year} {2022})},\ \Eprint {https://arxiv.org/abs/2202.10399}
  {arXiv:2202.10399 [cond-mat.quant-gas]} \BibitemShut {NoStop}%
\bibitem [{\citenamefont {Tajik}\ \emph
  {et~al.}(2023{\natexlab{b}})\citenamefont {Tajik}, \citenamefont {Gluza},
  \citenamefont {Sebe}, \citenamefont {Sch\"uttelkopf}, \citenamefont
  {Cataldini}, \citenamefont {Sabino}, \citenamefont {M{\o}ller}, \citenamefont
  {Ji}, \citenamefont {Erne}, \citenamefont {Guarnieri}, \citenamefont
  {Sotiriadis}, \citenamefont {Eisert},\ and\ \citenamefont
  {Schmiedmayer}}]{Tajik2023}%
  \BibitemOpen
  \bibfield  {author} {\bibinfo {author} {\bibfnamefont {M.}~\bibnamefont
  {Tajik}}, \bibinfo {author} {\bibfnamefont {M.}~\bibnamefont {Gluza}},
  \bibinfo {author} {\bibfnamefont {N.}~\bibnamefont {Sebe}}, \bibinfo {author}
  {\bibfnamefont {P.}~\bibnamefont {Sch\"uttelkopf}}, \bibinfo {author}
  {\bibfnamefont {F.}~\bibnamefont {Cataldini}}, \bibinfo {author}
  {\bibfnamefont {J.~a.}\ \bibnamefont {Sabino}}, \bibinfo {author}
  {\bibfnamefont {F.}~\bibnamefont {M{\o}ller}}, \bibinfo {author}
  {\bibfnamefont {S.}~\bibnamefont {Ji}}, \bibinfo {author} {\bibfnamefont
  {S.}~\bibnamefont {Erne}}, \bibinfo {author} {\bibfnamefont {G.}~\bibnamefont
  {Guarnieri}}, \bibinfo {author} {\bibfnamefont {S.}~\bibnamefont
  {Sotiriadis}}, \bibinfo {author} {\bibfnamefont {J.}~\bibnamefont {Eisert}},\
  and\ \bibinfo {author} {\bibfnamefont {J.}~\bibnamefont {Schmiedmayer}},\
  }\bibfield  {title} {\bibinfo {title} {Experimental observation of curved
  light-cones in a quantum field simulator},\ }\href
  {https://doi.org/10.1073/pnas.2301287120} {\bibfield  {journal} {\bibinfo
  {journal} {Proc. Natl. Acad. Sci. U.S.A.}\ }\textbf {\bibinfo {volume}
  {120}},\ \bibinfo {pages} {e2301287120} (\bibinfo {year}
  {2023}{\natexlab{b}})},\ \Eprint {https://arxiv.org/abs/2209.09132}
  {arXiv:2209.09132 [cond-mat.quant-gas]} \BibitemShut {NoStop}%
\bibitem [{\citenamefont {Garay}\ \emph {et~al.}(2000)\citenamefont {Garay},
  \citenamefont {Anglin}, \citenamefont {Cirac},\ and\ \citenamefont
  {Zoller}}]{PhysRevLett.85.4643}%
  \BibitemOpen
  \bibfield  {author} {\bibinfo {author} {\bibfnamefont {L.~J.}\ \bibnamefont
  {Garay}}, \bibinfo {author} {\bibfnamefont {J.~R.}\ \bibnamefont {Anglin}},
  \bibinfo {author} {\bibfnamefont {J.~I.}\ \bibnamefont {Cirac}},\ and\
  \bibinfo {author} {\bibfnamefont {P.}~\bibnamefont {Zoller}},\ }\bibfield
  {title} {\bibinfo {title} {Sonic analog of gravitational black holes in
  {Bose-Einstein} condensates},\ }\href
  {https://doi.org/10.1103/PhysRevLett.85.4643} {\bibfield  {journal} {\bibinfo
   {journal} {Phys. Rev. Lett.}\ }\textbf {\bibinfo {volume} {85}},\ \bibinfo
  {pages} {4643} (\bibinfo {year} {2000})}\BibitemShut {NoStop}%
\bibitem [{\citenamefont {Lahav}\ \emph {et~al.}(2010)\citenamefont {Lahav},
  \citenamefont {Itah}, \citenamefont {Blumkin}, \citenamefont {Gordon},
  \citenamefont {Rinott}, \citenamefont {Zayats},\ and\ \citenamefont
  {Steinhauer}}]{PhysRevLett.105.240401}%
  \BibitemOpen
  \bibfield  {author} {\bibinfo {author} {\bibfnamefont {O.}~\bibnamefont
  {Lahav}}, \bibinfo {author} {\bibfnamefont {A.}~\bibnamefont {Itah}},
  \bibinfo {author} {\bibfnamefont {A.}~\bibnamefont {Blumkin}}, \bibinfo
  {author} {\bibfnamefont {C.}~\bibnamefont {Gordon}}, \bibinfo {author}
  {\bibfnamefont {S.}~\bibnamefont {Rinott}}, \bibinfo {author} {\bibfnamefont
  {A.}~\bibnamefont {Zayats}},\ and\ \bibinfo {author} {\bibfnamefont
  {J.}~\bibnamefont {Steinhauer}},\ }\bibfield  {title} {\bibinfo {title}
  {Realization of a sonic black hole analog in a {Bose-Einstein} condensate},\
  }\href {https://doi.org/10.1103/PhysRevLett.105.240401} {\bibfield  {journal}
  {\bibinfo  {journal} {Phys. Rev. Lett.}\ }\textbf {\bibinfo {volume} {105}},\
  \bibinfo {pages} {240401} (\bibinfo {year} {2010})}\BibitemShut {NoStop}%
\bibitem [{\citenamefont {Steinhauer}(2016)}]{Steinhauer2016}%
  \BibitemOpen
  \bibfield  {author} {\bibinfo {author} {\bibfnamefont {J.}~\bibnamefont
  {Steinhauer}},\ }\bibfield  {title} {\bibinfo {title} {Observation of quantum
  {Hawking} radiation and its entanglement in an analogue black hole},\ }\href
  {https://doi.org/10.1038/nphys3863} {\bibfield  {journal} {\bibinfo
  {journal} {Nature Physics}\ }\textbf {\bibinfo {volume} {12}},\ \bibinfo
  {pages} {959} (\bibinfo {year} {2016})}\BibitemShut {NoStop}%
\bibitem [{\citenamefont {Navon}\ \emph {et~al.}(2021)\citenamefont {Navon},
  \citenamefont {Smith},\ and\ \citenamefont {Hadzibabic}}]{Navon2021}%
  \BibitemOpen
  \bibfield  {author} {\bibinfo {author} {\bibfnamefont {N.}~\bibnamefont
  {Navon}}, \bibinfo {author} {\bibfnamefont {R.~P.}\ \bibnamefont {Smith}},\
  and\ \bibinfo {author} {\bibfnamefont {Z.}~\bibnamefont {Hadzibabic}},\
  }\bibfield  {title} {\bibinfo {title} {Quantum gases in optical boxes},\
  }\href {https://doi.org/10.1038/s41567-021-01403-z} {\bibfield  {journal}
  {\bibinfo  {journal} {Nature Physics}\ }\textbf {\bibinfo {volume} {17}},\
  \bibinfo {pages} {1334} (\bibinfo {year} {2021})}\BibitemShut {NoStop}%
\bibitem [{\citenamefont {Gaunt}\ \emph {et~al.}(2013)\citenamefont {Gaunt},
  \citenamefont {Schmidutz}, \citenamefont {Gotlibovych}, \citenamefont
  {Smith},\ and\ \citenamefont {Hadzibabic}}]{Gaunt2013}%
  \BibitemOpen
  \bibfield  {author} {\bibinfo {author} {\bibfnamefont {A.~L.}\ \bibnamefont
  {Gaunt}}, \bibinfo {author} {\bibfnamefont {T.~F.}\ \bibnamefont
  {Schmidutz}}, \bibinfo {author} {\bibfnamefont {I.}~\bibnamefont
  {Gotlibovych}}, \bibinfo {author} {\bibfnamefont {R.~P.}\ \bibnamefont
  {Smith}},\ and\ \bibinfo {author} {\bibfnamefont {Z.}~\bibnamefont
  {Hadzibabic}},\ }\bibfield  {title} {\bibinfo {title} {Bose-einstein
  condensation of atoms in a uniform potential},\ }\href
  {https://doi.org/10.1103/PhysRevLett.110.200406} {\bibfield  {journal}
  {\bibinfo  {journal} {Phys. Rev. Lett.}\ }\textbf {\bibinfo {volume} {110}},\
  \bibinfo {pages} {200406} (\bibinfo {year} {2013})}\BibitemShut {NoStop}%
\bibitem [{\citenamefont {Jenkins}\ \emph
  {et~al.}(2024{\natexlab{b}})\citenamefont {Jenkins}, \citenamefont {Moss},
  \citenamefont {Billam}, \citenamefont {Hadzibabic}, \citenamefont {Peiris},\
  and\ \citenamefont {Pontzen}}]{Jenkins2024b}%
  \BibitemOpen
  \bibfield  {author} {\bibinfo {author} {\bibfnamefont {A.~C.}\ \bibnamefont
  {Jenkins}}, \bibinfo {author} {\bibfnamefont {I.~G.}\ \bibnamefont {Moss}},
  \bibinfo {author} {\bibfnamefont {T.~P.}\ \bibnamefont {Billam}}, \bibinfo
  {author} {\bibfnamefont {Z.}~\bibnamefont {Hadzibabic}}, \bibinfo {author}
  {\bibfnamefont {H.~V.}\ \bibnamefont {Peiris}},\ and\ \bibinfo {author}
  {\bibfnamefont {A.}~\bibnamefont {Pontzen}},\ }\bibfield  {title} {\bibinfo
  {title} {{Generalized cold-atom simulators for vacuum decay}},\ }\href
  {https://doi.org/10.1103/PhysRevA.110.L031301} {\bibfield  {journal}
  {\bibinfo  {journal} {Phys. Rev. A}\ }\textbf {\bibinfo {volume} {110}},\
  \bibinfo {pages} {L031301} (\bibinfo {year} {2024}{\natexlab{b}})},\ \Eprint
  {https://arxiv.org/abs/2311.02156} {arXiv:2311.02156 [cond-mat.quant-gas]}
  \BibitemShut {NoStop}%
\bibitem [{\citenamefont {Schweigler}\ \emph {et~al.}(2017)\citenamefont
  {Schweigler}, \citenamefont {Kasper}, \citenamefont {Erne}, \citenamefont
  {Mazets}, \citenamefont {Rauer}, \citenamefont {Cataldini}, \citenamefont
  {Langen}, \citenamefont {Gasenzer}, \citenamefont {Berges},\ and\
  \citenamefont {Schmiedmayer}}]{Schweigler2017}%
  \BibitemOpen
  \bibfield  {author} {\bibinfo {author} {\bibfnamefont {T.}~\bibnamefont
  {Schweigler}}, \bibinfo {author} {\bibfnamefont {V.}~\bibnamefont {Kasper}},
  \bibinfo {author} {\bibfnamefont {S.}~\bibnamefont {Erne}}, \bibinfo {author}
  {\bibfnamefont {I.}~\bibnamefont {Mazets}}, \bibinfo {author} {\bibfnamefont
  {B.}~\bibnamefont {Rauer}}, \bibinfo {author} {\bibfnamefont
  {F.}~\bibnamefont {Cataldini}}, \bibinfo {author} {\bibfnamefont
  {T.}~\bibnamefont {Langen}}, \bibinfo {author} {\bibfnamefont
  {T.}~\bibnamefont {Gasenzer}}, \bibinfo {author} {\bibfnamefont
  {J.}~\bibnamefont {Berges}},\ and\ \bibinfo {author} {\bibfnamefont
  {J.}~\bibnamefont {Schmiedmayer}},\ }\bibfield  {title} {\bibinfo {title} {On
  solving quantum many-body problems by experiment},\ }\href
  {https://doi.org/10.1038/nature22310} {\bibfield  {journal} {\bibinfo
  {journal} {Nature}\ }\textbf {\bibinfo {volume} {545}},\ \bibinfo {pages}
  {323} (\bibinfo {year} {2017})},\ \Eprint {https://arxiv.org/abs/1505.03126}
  {arXiv:1505.03126 [cond-mat.quant-gas]} \BibitemShut {NoStop}%
\bibitem [{\citenamefont {Lye}\ \emph {et~al.}(2003)\citenamefont {Lye},
  \citenamefont {Hope},\ and\ \citenamefont {Close}}]{PhysRevA.67.043609}%
  \BibitemOpen
  \bibfield  {author} {\bibinfo {author} {\bibfnamefont {J.~E.}\ \bibnamefont
  {Lye}}, \bibinfo {author} {\bibfnamefont {J.~J.}\ \bibnamefont {Hope}},\ and\
  \bibinfo {author} {\bibfnamefont {J.~D.}\ \bibnamefont {Close}},\ }\bibfield
  {title} {\bibinfo {title} {Nondestructive dynamic detectors for
  {Bose-Einstein} condensates},\ }\href
  {https://doi.org/10.1103/PhysRevA.67.043609} {\bibfield  {journal} {\bibinfo
  {journal} {Phys. Rev. A}\ }\textbf {\bibinfo {volume} {67}},\ \bibinfo
  {pages} {043609} (\bibinfo {year} {2003})}\BibitemShut {NoStop}%
\bibitem [{\citenamefont {Gooding}\ \emph {et~al.}(2020)\citenamefont
  {Gooding}, \citenamefont {Biermann}, \citenamefont {Erne}, \citenamefont
  {Louko}, \citenamefont {Unruh}, \citenamefont {Schmiedmayer},\ and\
  \citenamefont {Weinfurtner}}]{PhysRevLett.125.213603}%
  \BibitemOpen
  \bibfield  {author} {\bibinfo {author} {\bibfnamefont {C.}~\bibnamefont
  {Gooding}}, \bibinfo {author} {\bibfnamefont {S.}~\bibnamefont {Biermann}},
  \bibinfo {author} {\bibfnamefont {S.}~\bibnamefont {Erne}}, \bibinfo {author}
  {\bibfnamefont {J.}~\bibnamefont {Louko}}, \bibinfo {author} {\bibfnamefont
  {W.~G.}\ \bibnamefont {Unruh}}, \bibinfo {author} {\bibfnamefont
  {J.}~\bibnamefont {Schmiedmayer}},\ and\ \bibinfo {author} {\bibfnamefont
  {S.}~\bibnamefont {Weinfurtner}},\ }\bibfield  {title} {\bibinfo {title}
  {Interferometric {Unruh} detectors for {Bose-Einstein} condensates},\ }\href
  {https://doi.org/10.1103/PhysRevLett.125.213603} {\bibfield  {journal}
  {\bibinfo  {journal} {Phys. Rev. Lett.}\ }\textbf {\bibinfo {volume} {125}},\
  \bibinfo {pages} {213603} (\bibinfo {year} {2020})},\ \Eprint
  {https://arxiv.org/abs/2007.07160} {arXiv:2007.07160 [gr-qc]} \BibitemShut
  {NoStop}%
\bibitem [{\citenamefont {Fedichev}\ and\ \citenamefont
  {Fischer}(2004)}]{UweAndPetrObsDep2004}%
  \BibitemOpen
  \bibfield  {author} {\bibinfo {author} {\bibfnamefont {P.~O.}\ \bibnamefont
  {Fedichev}}\ and\ \bibinfo {author} {\bibfnamefont {U.~R.}\ \bibnamefont
  {Fischer}},\ }\bibfield  {title} {\bibinfo {title} {Observer dependence for
  the phonon content of the sound field living on the effective curved
  space-time background of a {Bose-Einstein} condensate},\ }\href
  {https://doi.org/10.1103/PhysRevD.69.064021} {\bibfield  {journal} {\bibinfo
  {journal} {Phys. Rev. D}\ }\textbf {\bibinfo {volume} {69}},\ \bibinfo
  {pages} {064021} (\bibinfo {year} {2004})},\ \Eprint
  {https://arxiv.org/abs/cond-mat/0307200} {arXiv:cond-mat/0307200 [cond-mat]}
  \BibitemShut {NoStop}%
\bibitem [{\citenamefont {Caves}(1980)}]{caves-PRL}%
  \BibitemOpen
  \bibfield  {author} {\bibinfo {author} {\bibfnamefont {C.~M.}\ \bibnamefont
  {Caves}},\ }\bibfield  {title} {\bibinfo {title} {Quantum-mechanical
  radiation-pressure fluctuations in an interferometer},\ }\href
  {https://doi.org/10.1103/PhysRevLett.45.75} {\bibfield  {journal} {\bibinfo
  {journal} {Phys. Rev. Lett.}\ }\textbf {\bibinfo {volume} {45}},\ \bibinfo
  {pages} {75} (\bibinfo {year} {1980})}\BibitemShut {NoStop}%
\bibitem [{\citenamefont {Caves}(1981)}]{second}%
  \BibitemOpen
  \bibfield  {author} {\bibinfo {author} {\bibfnamefont {C.~M.}\ \bibnamefont
  {Caves}},\ }\bibfield  {title} {\bibinfo {title} {Quantum-mechanical noise in
  an interferometer},\ }\href
  {https://doi.org/https://doi.org/10.1103/PhysRevD.23.1693} {\bibfield
  {journal} {\bibinfo  {journal} {Phys. Rev. D}\ }\textbf {\bibinfo {volume}
  {23}},\ \bibinfo {pages} {1693} (\bibinfo {year} {1981})}\BibitemShut
  {NoStop}%
\bibitem [{\citenamefont {Fulling}(1973)}]{Fulling}%
  \BibitemOpen
  \bibfield  {author} {\bibinfo {author} {\bibfnamefont {S.~A.}\ \bibnamefont
  {Fulling}},\ }\bibfield  {title} {\bibinfo {title} {{Nonuniqueness of
  canonical field quantization in Riemannian space-time}},\ }\href
  {https://doi.org/10.1103/PhysRevD.7.2850} {\bibfield  {journal} {\bibinfo
  {journal} {Phys. Rev. D}\ }\textbf {\bibinfo {volume} {7}},\ \bibinfo {pages}
  {2850} (\bibinfo {year} {1973})}\BibitemShut {NoStop}%
\bibitem [{\citenamefont {Davies}(1975)}]{Davies1975}%
  \BibitemOpen
  \bibfield  {author} {\bibinfo {author} {\bibfnamefont {P.~C.~W.}\
  \bibnamefont {Davies}},\ }\bibfield  {title} {\bibinfo {title} {{Scalar
  production in Schwarzschild and Rindler metrics}},\ }\href
  {https://doi.org/10.1088/0305-4470/8/4/022} {\bibfield  {journal} {\bibinfo
  {journal} {J. Phys. A: Math. Gen.}\ }\textbf {\bibinfo {volume} {8}},\
  \bibinfo {pages} {609} (\bibinfo {year} {1975})}\BibitemShut {NoStop}%
\bibitem [{\citenamefont {Unruh}(1976)}]{unruh76}%
  \BibitemOpen
  \bibfield  {author} {\bibinfo {author} {\bibfnamefont {W.~G.}\ \bibnamefont
  {Unruh}},\ }\bibfield  {title} {\bibinfo {title} {Notes on black hole
  evaporation},\ }\href
  {https://doi.org/https://doi.org/10.1103/PhysRevD.14.870} {\bibfield
  {journal} {\bibinfo  {journal} {Phys. Rev. D}\ }\textbf {\bibinfo {volume}
  {14}},\ \bibinfo {pages} {870} (\bibinfo {year} {1976})}\BibitemShut
  {NoStop}%
\bibitem [{\citenamefont {Fulling}\ and\ \citenamefont
  {Matsas}(2014)}]{Fulling:2014}%
  \BibitemOpen
  \bibfield  {author} {\bibinfo {author} {\bibfnamefont {S.~A.}\ \bibnamefont
  {Fulling}}\ and\ \bibinfo {author} {\bibfnamefont {G.~E.~A.}\ \bibnamefont
  {Matsas}},\ }\bibfield  {title} {\bibinfo {title} {{U}nruh effect},\ }\href
  {https://doi.org/10.4249/scholarpedia.31789} {\bibfield  {journal} {\bibinfo
  {journal} {Scholarpedia}\ }\textbf {\bibinfo {volume} {9}},\ \bibinfo {pages}
  {31789} (\bibinfo {year} {2014})}\BibitemShut {NoStop}%
\bibitem [{\citenamefont {Unruh}(2022)}]{Unruh:2022gso}%
  \BibitemOpen
  \bibfield  {author} {\bibinfo {author} {\bibfnamefont {W.~G.}\ \bibnamefont
  {Unruh}},\ }\bibfield  {title} {\bibinfo {title} {Black holes, acceleration
  temperature and low temperature analog experiments},\ }\href
  {https://doi.org/10.1007/s10909-021-02662-z} {\bibfield  {journal} {\bibinfo
  {journal} {J. Low Temp. Phys.}\ }\textbf {\bibinfo {volume} {208}},\ \bibinfo
  {pages} {196} (\bibinfo {year} {2022})}\BibitemShut {NoStop}%
\bibitem [{sup()}]{suppmat}%
  \BibitemOpen
  \href@noop {} {\bibinfo {title} {See supplemental material.}}\BibitemShut
  {Stop}%
\bibitem [{\citenamefont {Khalili}\ \emph {et~al.}(2012)\citenamefont
  {Khalili}, \citenamefont {Miao}, \citenamefont {Yang}, \citenamefont
  {Safavi-Naeini}, \citenamefont {Painter},\ and\ \citenamefont
  {Chen}}]{Khalili2012}%
  \BibitemOpen
  \bibfield  {author} {\bibinfo {author} {\bibfnamefont {F.~Y.}\ \bibnamefont
  {Khalili}}, \bibinfo {author} {\bibfnamefont {H.}~\bibnamefont {Miao}},
  \bibinfo {author} {\bibfnamefont {H.}~\bibnamefont {Yang}}, \bibinfo {author}
  {\bibfnamefont {A.~H.}\ \bibnamefont {Safavi-Naeini}}, \bibinfo {author}
  {\bibfnamefont {O.}~\bibnamefont {Painter}},\ and\ \bibinfo {author}
  {\bibfnamefont {Y.}~\bibnamefont {Chen}},\ }\bibfield  {title} {\bibinfo
  {title} {Quantum back-action in measurements of zero-point mechanical
  oscillations},\ }\href {https://doi.org/10.1103/physreva.86.033840}
  {\bibfield  {journal} {\bibinfo  {journal} {Phys. Rev. A}\ }\textbf {\bibinfo
  {volume} {86}},\ \bibinfo {pages} {033840} (\bibinfo {year}
  {2012})}\BibitemShut {NoStop}%
\bibitem [{\citenamefont {{DeWitt}}(1979)}]{DeWitt1979}%
  \BibitemOpen
  \bibfield  {author} {\bibinfo {author} {\bibfnamefont {B.~S.}\ \bibnamefont
  {{DeWitt}}},\ }\bibfield  {title} {\bibinfo {title} {{Quantum gravity: The
  new synthesis}},\ }in\ \href@noop {} {\emph {\bibinfo {booktitle} {General
  Relativity: An Einstein Centenary Survey}}},\ \bibinfo {editor} {edited by\
  \bibinfo {editor} {\bibfnamefont {S.~W.}\ \bibnamefont {{Hawking}}}\ and\
  \bibinfo {editor} {\bibfnamefont {W.}~\bibnamefont {{Israel}}}}\ (\bibinfo
  {publisher} {Cambridge University Press, Cambridge, England},\ \bibinfo
  {year} {1979})\ pp.\ \bibinfo {pages} {680--745}\BibitemShut {NoStop}%
\bibitem [{\citenamefont {Biermann}\ \emph {et~al.}(2020)\citenamefont
  {Biermann}, \citenamefont {Erne}, \citenamefont {Gooding}, \citenamefont
  {Louko}, \citenamefont {Schmiedmayer}, \citenamefont {Unruh},\ and\
  \citenamefont {Weinfurtner}}]{PhysRevD.102.085006}%
  \BibitemOpen
  \bibfield  {author} {\bibinfo {author} {\bibfnamefont {S.}~\bibnamefont
  {Biermann}}, \bibinfo {author} {\bibfnamefont {S.}~\bibnamefont {Erne}},
  \bibinfo {author} {\bibfnamefont {C.}~\bibnamefont {Gooding}}, \bibinfo
  {author} {\bibfnamefont {J.}~\bibnamefont {Louko}}, \bibinfo {author}
  {\bibfnamefont {J.}~\bibnamefont {Schmiedmayer}}, \bibinfo {author}
  {\bibfnamefont {W.~G.}\ \bibnamefont {Unruh}},\ and\ \bibinfo {author}
  {\bibfnamefont {S.}~\bibnamefont {Weinfurtner}},\ }\bibfield  {title}
  {\bibinfo {title} {Unruh and analogue {Unruh} temperatures for circular
  motion in $3+1$ and $2+1$ dimensions},\ }\href
  {https://doi.org/10.1103/PhysRevD.102.085006} {\bibfield  {journal} {\bibinfo
   {journal} {Phys. Rev. D}\ }\textbf {\bibinfo {volume} {102}},\ \bibinfo
  {pages} {085006} (\bibinfo {year} {2020})},\ \Eprint
  {https://arxiv.org/abs/2007.09523} {arXiv:2007.09523 [gr-qc]} \BibitemShut
  {NoStop}%
\bibitem [{\citenamefont {Bowen}\ and\ \citenamefont
  {Milburn}(2015)}]{bowenmilburn}%
  \BibitemOpen
  \bibfield  {author} {\bibinfo {author} {\bibfnamefont {W.~P.}\ \bibnamefont
  {Bowen}}\ and\ \bibinfo {author} {\bibfnamefont {G.~J.}\ \bibnamefont
  {Milburn}},\ }\href {https://doi.org/https://doi.org/10.1201/b19379} {\emph
  {\bibinfo {title} {Quantum Optomechanics}}}\ (\bibinfo  {publisher} {CRC
  Press, Boca Raton, Florida},\ \bibinfo {year} {2015})\BibitemShut {NoStop}%
\bibitem [{\citenamefont {Glauber}(1963)}]{glauber63}%
  \BibitemOpen
  \bibfield  {author} {\bibinfo {author} {\bibfnamefont {R.~J.}\ \bibnamefont
  {Glauber}},\ }\bibfield  {title} {\bibinfo {title} {The quantum theory of
  optical coherence},\ }\href
  {https://doi.org/https://doi.org/10.1103/PhysRev.130.2529} {\bibfield
  {journal} {\bibinfo  {journal} {Phys. Rev.}\ }\textbf {\bibinfo {volume}
  {130}},\ \bibinfo {pages} {2529} (\bibinfo {year} {1963})}\BibitemShut
  {NoStop}%
\bibitem [{\citenamefont {Ou}\ \emph {et~al.}(1987)\citenamefont {Ou},
  \citenamefont {Hong},\ and\ \citenamefont {Mandel}}]{Ou1987}%
  \BibitemOpen
  \bibfield  {author} {\bibinfo {author} {\bibfnamefont {Z.~Y.}\ \bibnamefont
  {Ou}}, \bibinfo {author} {\bibfnamefont {C.~K.}\ \bibnamefont {Hong}},\ and\
  \bibinfo {author} {\bibfnamefont {L.}~\bibnamefont {Mandel}},\ }\bibfield
  {title} {\bibinfo {title} {Coherence properties of squeezed light and the
  degree of squeezing},\ }\href {https://doi.org/10.1364/josab.4.001574}
  {\bibfield  {journal} {\bibinfo  {journal} {J. Opt. Soc. Am. B}\ }\textbf
  {\bibinfo {volume} {4}},\ \bibinfo {pages} {1574} (\bibinfo {year}
  {1987})}\BibitemShut {NoStop}%
\bibitem [{\citenamefont {Yuen}\ and\ \citenamefont {Chan}(1983)}]{Yuen1983}%
  \BibitemOpen
  \bibfield  {author} {\bibinfo {author} {\bibfnamefont {H.~P.}\ \bibnamefont
  {Yuen}}\ and\ \bibinfo {author} {\bibfnamefont {V.~W.~S.}\ \bibnamefont
  {Chan}},\ }\bibfield  {title} {\bibinfo {title} {Noise in homodyne and
  heterodyne detection},\ }\href {https://doi.org/10.1364/ol.8.000177}
  {\bibfield  {journal} {\bibinfo  {journal} {Optics Letters}\ }\textbf
  {\bibinfo {volume} {8}},\ \bibinfo {pages} {177} (\bibinfo {year}
  {1983})}\BibitemShut {NoStop}%
\bibitem [{\citenamefont {Shapiro}(1985)}]{quantumnoise1985}%
  \BibitemOpen
  \bibfield  {author} {\bibinfo {author} {\bibfnamefont {J.}~\bibnamefont
  {Shapiro}},\ }\bibfield  {title} {\bibinfo {title} {Quantum noise and excess
  noise in optical homodyne and heterodyne receivers},\ }\href
  {https://doi.org/10.1109/JQE.1985.1072640} {\bibfield  {journal} {\bibinfo
  {journal} {IEEE Journal of Quantum Electronics}\ }\textbf {\bibinfo {volume}
  {21}},\ \bibinfo {pages} {237} (\bibinfo {year} {1985})}\BibitemShut
  {NoStop}%
\bibitem [{\citenamefont {Collett}\ \emph {et~al.}(1987)\citenamefont
  {Collett}, \citenamefont {Loudon},\ and\ \citenamefont
  {Gardiner}}]{Collett1987}%
  \BibitemOpen
  \bibfield  {author} {\bibinfo {author} {\bibfnamefont {M.}~\bibnamefont
  {Collett}}, \bibinfo {author} {\bibfnamefont {R.}~\bibnamefont {Loudon}},\
  and\ \bibinfo {author} {\bibfnamefont {C.}~\bibnamefont {Gardiner}},\
  }\bibfield  {title} {\bibinfo {title} {Quantum theory of optical homodyne and
  heterodyne detection},\ }\href {https://doi.org/10.1080/09500348714550811}
  {\bibfield  {journal} {\bibinfo  {journal} {J. Mod. Optics}\ }\textbf
  {\bibinfo {volume} {34}},\ \bibinfo {pages} {881} (\bibinfo {year}
  {1987})}\BibitemShut {NoStop}%
\bibitem [{\citenamefont {Li}\ \emph {et~al.}(1999)\citenamefont {Li},
  \citenamefont {Guzun},\ and\ \citenamefont {Xiao}}]{PhysRevLett.82.5225}%
  \BibitemOpen
  \bibfield  {author} {\bibinfo {author} {\bibfnamefont {Y.-q.}\ \bibnamefont
  {Li}}, \bibinfo {author} {\bibfnamefont {D.}~\bibnamefont {Guzun}},\ and\
  \bibinfo {author} {\bibfnamefont {M.}~\bibnamefont {Xiao}},\ }\bibfield
  {title} {\bibinfo {title} {Sub-shot-noise-limited optical heterodyne
  detection using an amplitude-squeezed local oscillator},\ }\href
  {https://doi.org/10.1103/PhysRevLett.82.5225} {\bibfield  {journal} {\bibinfo
   {journal} {Phys. Rev. Lett.}\ }\textbf {\bibinfo {volume} {82}},\ \bibinfo
  {pages} {5225} (\bibinfo {year} {1999})}\BibitemShut {NoStop}%
\bibitem [{\citenamefont {Eberle}\ \emph {et~al.}(2013)\citenamefont {Eberle},
  \citenamefont {H\"{a}ndchen},\ and\ \citenamefont {Schnabel}}]{Eberle2013}%
  \BibitemOpen
  \bibfield  {author} {\bibinfo {author} {\bibfnamefont {T.}~\bibnamefont
  {Eberle}}, \bibinfo {author} {\bibfnamefont {V.}~\bibnamefont
  {H\"{a}ndchen}},\ and\ \bibinfo {author} {\bibfnamefont {R.}~\bibnamefont
  {Schnabel}},\ }\bibfield  {title} {\bibinfo {title} {Stable control of 10 db
  two-mode squeezed vacuum states of light},\ }\href
  {https://doi.org/10.1364/oe.21.011546} {\bibfield  {journal} {\bibinfo
  {journal} {Optics Express}\ }\textbf {\bibinfo {volume} {21}},\ \bibinfo
  {pages} {11546} (\bibinfo {year} {2013})}\BibitemShut {NoStop}%
\bibitem [{\citenamefont {Lueghamer}\ \emph {et~al.}(2025)\citenamefont
  {Lueghamer}, \citenamefont {Nimmrichter}, \citenamefont {Conrad-Billroth},
  \citenamefont {Juffmann},\ and\ \citenamefont {Pr\"{u}fer}}]{Lueghamer2025}%
  \BibitemOpen
  \bibfield  {author} {\bibinfo {author} {\bibfnamefont {O.}~\bibnamefont
  {Lueghamer}}, \bibinfo {author} {\bibfnamefont {S.}~\bibnamefont
  {Nimmrichter}}, \bibinfo {author} {\bibfnamefont {C.}~\bibnamefont
  {Conrad-Billroth}}, \bibinfo {author} {\bibfnamefont {T.}~\bibnamefont
  {Juffmann}},\ and\ \bibinfo {author} {\bibfnamefont {M.}~\bibnamefont
  {Pr\"{u}fer}},\ }\bibfield  {title} {\bibinfo {title} {Cavity-enhanced
  continuous-wave microscopy with potentially unstable cavity length},\ }\href
  {https://doi.org/10.1038/s41598-025-13589-w} {\bibfield  {journal} {\bibinfo
  {journal} {Scientific Reports}\ }\textbf {\bibinfo {volume} {15}},\ \bibinfo
  {pages} {27676} (\bibinfo {year} {2025})}\BibitemShut {NoStop}%
\bibitem [{\citenamefont {Letaw}(1981)}]{LetawStationary}%
  \BibitemOpen
  \bibfield  {author} {\bibinfo {author} {\bibfnamefont {J.~R.}\ \bibnamefont
  {Letaw}},\ }\bibfield  {title} {\bibinfo {title} {{Stationary world lines and
  the vacuum excitation of noninertial detectors}},\ }\href
  {https://doi.org/10.1103/PhysRevD.23.1709} {\bibfield  {journal} {\bibinfo
  {journal} {Phys. Rev. D}\ }\textbf {\bibinfo {volume} {23}},\ \bibinfo
  {pages} {1709} (\bibinfo {year} {1981})}\BibitemShut {NoStop}%
\bibitem [{\citenamefont {Bunney}(2024)}]{Bunney:2023mkh}%
  \BibitemOpen
  \bibfield  {author} {\bibinfo {author} {\bibfnamefont {C.~R.~D.}\
  \bibnamefont {Bunney}},\ }\bibfield  {title} {\bibinfo {title} {{Stationary
  trajectories in Minkowski spacetimes}},\ }\href
  {https://doi.org/10.1063/5.0205471} {\bibfield  {journal} {\bibinfo
  {journal} {J. Math. Phys.}\ }\textbf {\bibinfo {volume} {65}},\ \bibinfo
  {pages} {052501} (\bibinfo {year} {2024})},\ \Eprint
  {https://arxiv.org/abs/2310.04359} {arXiv:2310.04359 [gr-qc]} \BibitemShut
  {NoStop}%
\bibitem [{\citenamefont {Letaw}\ and\ \citenamefont
  {Pfautsch}(1981)}]{Letaw:1980ik}%
  \BibitemOpen
  \bibfield  {author} {\bibinfo {author} {\bibfnamefont {J.~R.}\ \bibnamefont
  {Letaw}}\ and\ \bibinfo {author} {\bibfnamefont {J.~D.}\ \bibnamefont
  {Pfautsch}},\ }\bibfield  {title} {\bibinfo {title} {{The Quantized Scalar
  Field in the Stationary Coordinate Systems of Flat Space-time}},\ }\href
  {https://doi.org/10.1103/PhysRevD.24.1491} {\bibfield  {journal} {\bibinfo
  {journal} {Phys. Rev. D}\ }\textbf {\bibinfo {volume} {24}},\ \bibinfo
  {pages} {1491} (\bibinfo {year} {1981})}\BibitemShut {NoStop}%
\bibitem [{\citenamefont {Good}\ \emph {et~al.}(2020)\citenamefont {Good},
  \citenamefont {Ju\'arez-Aubry}, \citenamefont {Moustos},\ and\ \citenamefont
  {Temirkhan}}]{Good:2020hav}%
  \BibitemOpen
  \bibfield  {author} {\bibinfo {author} {\bibfnamefont {M.}~\bibnamefont
  {Good}}, \bibinfo {author} {\bibfnamefont {B.~A.}\ \bibnamefont
  {Ju\'arez-Aubry}}, \bibinfo {author} {\bibfnamefont {D.}~\bibnamefont
  {Moustos}},\ and\ \bibinfo {author} {\bibfnamefont {M.}~\bibnamefont
  {Temirkhan}},\ }\bibfield  {title} {\bibinfo {title} {{Unruh-like effects:
  Effective temperatures along stationary worldlines}},\ }\href
  {https://doi.org/10.1007/JHEP06(2020)059} {\bibfield  {journal} {\bibinfo
  {journal} {JHEP}\ }\textbf {\bibinfo {volume} {06}},\ \bibinfo {pages}
  {059}},\ \Eprint {https://arxiv.org/abs/2004.08225} {arXiv:2004.08225
  [gr-qc]} \BibitemShut {NoStop}%
\bibitem [{\citenamefont {Bunney}\ and\ \citenamefont
  {Louko}(2023)}]{Bunney:2023vyj}%
  \BibitemOpen
  \bibfield  {author} {\bibinfo {author} {\bibfnamefont {C.~R.~D.}\
  \bibnamefont {Bunney}}\ and\ \bibinfo {author} {\bibfnamefont
  {J.}~\bibnamefont {Louko}},\ }\bibfield  {title} {\bibinfo {title} {{Circular
  motion analogue Unruh effect in a thermal bath: robbing from the rich and
  giving to the poor}},\ }\href {https://doi.org/10.1088/1361-6382/acde3b}
  {\bibfield  {journal} {\bibinfo  {journal} {Class. Quant. Grav.}\ }\textbf
  {\bibinfo {volume} {40}},\ \bibinfo {pages} {155001} (\bibinfo {year}
  {2023})},\ \Eprint {https://arxiv.org/abs/2303.12690} {arXiv:2303.12690
  [gr-qc]} \BibitemShut {NoStop}%
\bibitem [{\citenamefont {Bunney}\ \emph {et~al.}(2024)\citenamefont {Bunney},
  \citenamefont {Barroso}, \citenamefont {Biermann}, \citenamefont
  {Geelmuyden}, \citenamefont {Gooding}, \citenamefont {Ithier}, \citenamefont
  {Rojas}, \citenamefont {Louko},\ and\ \citenamefont
  {Weinfurtner}}]{https://doi.org/10.48550/arxiv.2302.12023}%
  \BibitemOpen
  \bibfield  {author} {\bibinfo {author} {\bibfnamefont {C.~R.~D.}\
  \bibnamefont {Bunney}}, \bibinfo {author} {\bibfnamefont {V.~S.}\
  \bibnamefont {Barroso}}, \bibinfo {author} {\bibfnamefont {S.}~\bibnamefont
  {Biermann}}, \bibinfo {author} {\bibfnamefont {A.}~\bibnamefont
  {Geelmuyden}}, \bibinfo {author} {\bibfnamefont {C.}~\bibnamefont {Gooding}},
  \bibinfo {author} {\bibfnamefont {G.}~\bibnamefont {Ithier}}, \bibinfo
  {author} {\bibfnamefont {X.}~\bibnamefont {Rojas}}, \bibinfo {author}
  {\bibfnamefont {J.}~\bibnamefont {Louko}},\ and\ \bibinfo {author}
  {\bibfnamefont {S.}~\bibnamefont {Weinfurtner}},\ }\bibfield  {title}
  {\bibinfo {title} {{Third sound detectors in accelerated motion}},\ }\href
  {https://doi.org/10.1088/1367-2630/ad5758} {\bibfield  {journal} {\bibinfo
  {journal} {New J. Phys.}\ }\textbf {\bibinfo {volume} {26}},\ \bibinfo
  {pages} {065001} (\bibinfo {year} {2024})},\ \Eprint
  {https://arxiv.org/abs/2302.12023} {arXiv:2302.12023 [gr-qc]} \BibitemShut
  {NoStop}%
\bibitem [{\citenamefont {Unruh}(1998)}]{Unruh:1998gq}%
  \BibitemOpen
  \bibfield  {author} {\bibinfo {author} {\bibfnamefont {W.~G.}\ \bibnamefont
  {Unruh}},\ }\bibfield  {title} {\bibinfo {title} {{Acceleration radiation for
  orbiting electrons}},\ }\href {https://doi.org/10.1016/S0370-1573(98)00068-4}
  {\bibfield  {journal} {\bibinfo  {journal} {Phys. Rept.}\ }\textbf {\bibinfo
  {volume} {307}},\ \bibinfo {pages} {163} (\bibinfo {year} {1998})},\ \Eprint
  {https://arxiv.org/abs/hep-th/9804158} {arXiv:hep-th/9804158} \BibitemShut
  {NoStop}%
\bibitem [{\citenamefont {Brahms}\ \emph {et~al.}(2012)\citenamefont {Brahms},
  \citenamefont {Botter}, \citenamefont {Schreppler}, \citenamefont {Brooks},\
  and\ \citenamefont {Stamper-Kurn}}]{PhysRevLett.108.133601}%
  \BibitemOpen
  \bibfield  {author} {\bibinfo {author} {\bibfnamefont {N.}~\bibnamefont
  {Brahms}}, \bibinfo {author} {\bibfnamefont {T.}~\bibnamefont {Botter}},
  \bibinfo {author} {\bibfnamefont {S.}~\bibnamefont {Schreppler}}, \bibinfo
  {author} {\bibfnamefont {D.~W.~C.}\ \bibnamefont {Brooks}},\ and\ \bibinfo
  {author} {\bibfnamefont {D.~M.}\ \bibnamefont {Stamper-Kurn}},\ }\bibfield
  {title} {\bibinfo {title} {Optical detection of the quantization of
  collective atomic motion},\ }\href
  {https://doi.org/10.1103/PhysRevLett.108.133601} {\bibfield  {journal}
  {\bibinfo  {journal} {Phys. Rev. Lett.}\ }\textbf {\bibinfo {volume} {108}},\
  \bibinfo {pages} {133601} (\bibinfo {year} {2012})}\BibitemShut {NoStop}%
\bibitem [{\citenamefont {Shkarin}\ \emph {et~al.}(2019)\citenamefont
  {Shkarin}, \citenamefont {Kashkanova}, \citenamefont {Brown}, \citenamefont
  {Garcia}, \citenamefont {Ott}, \citenamefont {Reichel},\ and\ \citenamefont
  {Harris}}]{PhysRevLett.122.153601}%
  \BibitemOpen
  \bibfield  {author} {\bibinfo {author} {\bibfnamefont {A.~B.}\ \bibnamefont
  {Shkarin}}, \bibinfo {author} {\bibfnamefont {A.~D.}\ \bibnamefont
  {Kashkanova}}, \bibinfo {author} {\bibfnamefont {C.~D.}\ \bibnamefont
  {Brown}}, \bibinfo {author} {\bibfnamefont {S.}~\bibnamefont {Garcia}},
  \bibinfo {author} {\bibfnamefont {K.}~\bibnamefont {Ott}}, \bibinfo {author}
  {\bibfnamefont {J.}~\bibnamefont {Reichel}},\ and\ \bibinfo {author}
  {\bibfnamefont {J.~G.~E.}\ \bibnamefont {Harris}},\ }\bibfield  {title}
  {\bibinfo {title} {Quantum optomechanics in a liquid},\ }\href
  {https://doi.org/10.1103/PhysRevLett.122.153601} {\bibfield  {journal}
  {\bibinfo  {journal} {Phys. Rev. Lett.}\ }\textbf {\bibinfo {volume} {122}},\
  \bibinfo {pages} {153601} (\bibinfo {year} {2019})}\BibitemShut {NoStop}%
\bibitem [{\citenamefont {Gupta}\ \emph {et~al.}(2024)\citenamefont {Gupta},
  \citenamefont {Kumar}, \citenamefont {Kanamoto}, \citenamefont
  {Bhattacharya},\ and\ \citenamefont {Dhar}}]{Gupta:2024mhs}%
  \BibitemOpen
  \bibfield  {author} {\bibinfo {author} {\bibfnamefont {R.}~\bibnamefont
  {Gupta}}, \bibinfo {author} {\bibfnamefont {P.}~\bibnamefont {Kumar}},
  \bibinfo {author} {\bibfnamefont {R.}~\bibnamefont {Kanamoto}}, \bibinfo
  {author} {\bibfnamefont {M.}~\bibnamefont {Bhattacharya}},\ and\ \bibinfo
  {author} {\bibfnamefont {H.~S.}\ \bibnamefont {Dhar}},\ }\bibfield  {title}
  {\bibinfo {title} {{Sensing atomic superfluid rotation beyond the standard
  quantum limit}},\ }\href {https://doi.org/10.1103/PhysRevA.110.053514}
  {\bibfield  {journal} {\bibinfo  {journal} {Phys. Rev. A}\ }\textbf {\bibinfo
  {volume} {110}},\ \bibinfo {pages} {053514} (\bibinfo {year} {2024})},\
  \Eprint {https://arxiv.org/abs/2402.19123} {arXiv:2402.19123 [quant-ph]}
  \BibitemShut {NoStop}%
\bibitem [{\citenamefont {Birrell}\ and\ \citenamefont {Davies}(1982)}]{BD}%
  \BibitemOpen
  \bibfield  {author} {\bibinfo {author} {\bibfnamefont {N.~D.}\ \bibnamefont
  {Birrell}}\ and\ \bibinfo {author} {\bibfnamefont {P.~C.~W.}\ \bibnamefont
  {Davies}},\ }\href {https://doi.org/https://doi.org/10.1017/CBO9780511622632}
  {\emph {\bibinfo {title} {Quantum fields in curved space}}}\ (\bibinfo
  {publisher} {Cambridge University Press},\ \bibinfo {year}
  {1982})\BibitemShut {NoStop}%
\end{thebibliography}%

\clearpage \raggedbottom

\begin{appendix}

\numberwithin{equation}{section}

\begin{center}  
{\large\bf End Matter} 
\end{center}

\section{Effective field theory}

In this Appendix, we use the effective field theory for the BEC field $\phi$ and the laser-BEC interaction to derive the interaction coupling $\varepsilon$ appearing in the definition of the dimensionless effective coupling parameter $\mu=-\varepsilon\alpha/\sqrt{2}$.

We describe the $2d$ BEC in terms of the two-dimensional Gross-Pitaevskii field $\Phi(t, \bm{x})$, with an isotropic potential $V(r)$:
\begin{align}
  i \partial_t \Phi=-\frac{1}{2 m} \nabla^2 \Phi+V\,\Phi+g_{2d} \Phi|\Phi|^2  \,.
\end{align}
Now consider small fluctuations $\Psi$ of the field $\Phi$ about a background $\Phi_0$ (i.e. $\Phi=\Phi_0+\Psi$). Defining the background BEC density $\rho_0=\Phi_0^2$ and assuming $\Phi_0$ is static, real-valued, and extends to infinity in $r$, we can then take $V+g_{2d} \rho_0=0$ and express the linearized equations of motion for $\Psi$ as
\begin{equation}
\begin{aligned}
 i \partial_{t} \Psi=-\frac{1}{2 m} \nabla^2 \Psi+g_{2d} \rho_0\left[\Psi+\Psi^\dagger\right]\, ,
\end{aligned}
\end{equation}
and similarly for $\Psi^\dagger$. The corresponding Lagrangian is
\begin{equation}
L_{\Psi}=\int d^2 x\, \left[i \Psi^{\dagger} \partial_{t} \Psi- \frac{1}{2 m}|\nabla \Psi|^2-\frac{g_{2d} \rho_0}{2}\left(\Psi+\Psi^{\dagger}\right)^2\right].
\end{equation} 

Writing the field $\Psi$ in terms of real and imaginary parts ($\Psi=\Psi_R+i \Psi_I$), the dynamics can alternatively be expressed as
\begin{equation}
\partial_t \Psi_R-\frac{1}{2 m} \nabla^2 \Psi_I=0
\end{equation}
and 
\begin{equation}
\partial_{t} \Psi_I+\frac{1}{2 m} \nabla^2 \Psi_R-2 g_{2d} \rho_0 \Psi_R=0\,.
\end{equation}
Combining these equations to eliminate $\Psi_I$ and taking the long-wavelength limit then leads to a Klein-Gordon equation for $\Psi_R$, with propagation speed $c_s=\sqrt{g_{2d} \rho_0/m}$. From the corresponding Lagrangian
\begin{equation}\label{LBEC}
    L_{\Psi_R}=\int d^2 x\left[\frac{1}{c_s^2}\left(\frac{\partial_t
    \Psi_R}{\sqrt{m}}\right)^2-\left(\frac{\nabla\Psi_R}{\sqrt{m}}\right)^2\right]\, ,
\end{equation}
we identify $\phi(t,\bm{x})=\Psi_R(t,\bm{x})/\sqrt{m}$ as our effective $(2+1)$-dimensional Klein-Gordon field. 

The first-order laser-BEC interaction \eqref{eq:firstorder} is characterised by the interaction Lagrangian~\cite{PhysRevLett.125.213603}
\begin{equation}\label{interaction}
L_{\text{int}}=-\frac{A_0^2}{2} \hat{\alpha}_R\omega_0 \delta \rho \partial_t \psi\, ,
\end{equation}
evaluated along the interaction trajectory, where $\delta \rho = 2\sqrt{\rho_0}\Psi_R$, $\hat{\alpha}_R$ is the real part of the polarisability, $\psi(t,z)$ is the effective laser phase, and $A_0\sim \alpha$ is the laser amplitude (assumed to be unaltered by the interaction). The (complex) polarisability $\hat{\alpha}=\hat{\alpha}_R+\mathrm{i}\hat{\alpha}_I$ is given explicitly by 
\begin{equation}\label{eqn:: polarizability}
\hat{\alpha}(\omega)=-\frac{24 \pi^2}{\omega_r^3} \frac{1}{\delta_0}
\! \left(1 - \frac{\I}{\delta_0} \right)
\, ,
\end{equation} 
where $\omega_r=\omega_0$ is the atomic resonance frequency and $\delta_0$ is the detuning from resonance (in units of atomic half-linewidths). The laser phase obeys the effective Lagrangian
\begin{equation}
 L_{\text {EM}}=\frac{A_0^2}{4}\int dz\, \left[\frac{1}{c_\text{eff}^2}\left(\partial_t{\psi}\right)^2-\left(\partial_z \psi\right)^2 \right]\ ,
\end{equation}
where $c_\text{eff}$ is the effective speed of light. Subject to the interaction \eqref{interaction}, the input laser phase $\psi_0(t,z)$ becomes~\cite{PhysRevLett.125.213603}
\begin{equation}
\psi=\psi_0\pm\frac{\varepsilon}{2} \phi=\psi_0\pm\frac{|\hat{\alpha}_R|\omega_0}{2}\,\delta \rho\, ,
\end{equation}where the $\pm$ sign corresponds to the opposite detuning in each sideband. Hence, we can identify $\varepsilon=2|\hat{\alpha}_R|\omega_0 \sqrt{m\, \rho_0}$, from which the dimensionless effective coupling parameter $\mu$ is found to be
\begin{equation}
\mu=-\frac{\varepsilon \alpha}{\sqrt{2}}=-|\hat{\alpha}_R|\omega_0\sqrt{2\,m\,\rho_0}\alpha \, .
\end{equation}

\section{Backaction from the Bogoliubov transformation}

In this appendix we give the Bogoliubov transformation through which the first-order interaction \eqref{eq:firstorder}
leads to the post-interaction operators~\eqref{delta_atilde}, 
including the crucial quadratic backaction term~\eqref{eq:baa}. We follow the procedure introduced in~\cite{Unruh:2022gso}, adapting it to our notation. 

In the notation of the main text, we introduce the new pre-interaction mode operators 
\begin{subequations}
\label{eqn:: common-and-difference-modes}
\begin{align}
z_\nu~=~&\frac{1}{\sqrt{2}}\left(\delta a_+[\nu]+\delta a_-[\nu]\right)\,,\\
Z_\nu~=~&\frac{1}{\sqrt{2}}\left(\delta a_+[\nu]-\delta a_-[\nu]\right)\,.\label{eqn:: common difference mode}  
\end{align}\end{subequations}
We call $z_\nu$ the common-mode operator and $Z_\nu$ the difference-mode operator. 
The transformation \eqref{eqn:: common-and-difference-modes} can be inverted to give \begin{align}
\delta a_\pm[\nu]=\frac{1}{\sqrt{2}}\left(z_\nu\pm Z_\nu\right) \,. 
\label{eqn:: inversi-common-and-difference-modes}
\end{align}

Next, as the operators $D_\nu$ in the BEC field mode decomposition \eqref{eqn:: phi decomp} 
are annihilation operators for $\nu>0$ and creation operators for $\nu<0$, satisfying 
$D_{-\nu}=D_\nu^\dagger$, 
we from now on assume $\nu\in (0,\Delta)$ and write 
\begin{align} 
Z_\nu = X_\nu, \ \ Z_{-\nu} = Y_\nu \ \ \ (\nu>0) \,. 
\label{eq:XY-def}
\end{align}
The full set of independent pre-interaction operators is then $(z_\nu, z_{-\nu}, X_\nu, Y_\nu, D_\nu)$, with $\nu\in (0,\Delta)$. 

For the corresponding post-interaction operators, \eqref{eq:firstorder} and \eqref{eqn:: phi decomp} give 
\begin{equation}\label{eqn:: first-order}
\tilde{z}_{\pm\nu}=z_{\pm\nu}\,, 
\quad
\tilde{X}_\nu=X_\nu+\I\mu D_\nu\,,
\quad
\tilde{Y}_\nu=Y_\nu+\I\mu D_\nu^\dagger\, ,
\end{equation}
where $\mu=-\varepsilon\alpha/\sqrt{2}<0$ is the dimensionless laser-BEC coupling parameter. The signal from the BEC is hence carried entirely in the difference-mode operators $X_\nu$ and~$Y_\nu$, whereas the common-mode operators $z_{\pm\nu}$ remain unaffected by the interaction with the BEC\null. 

We now complete the linear order transformation \eqref{eqn:: first-order} into a   
nonperturbative Bogoliubov transformation that includes quadratic terms in~$\mu$. 

We introduce the triple $u(\nu)=(X_\nu,Y_\nu,D_\nu)$, whose commutators are $[u_i(\nu),u_j^\dagger(\nu')]=\delta_{ij}\, 2\pi \delta(\nu-\nu')$ and $[u_i(\nu),u_j(\nu')]=0$, 
and a similar triple $\tilde{u}$ for the post-interaction operators, with similar commutators. 
We look for a nonperturbative Bogoliubov transformation
\begin{equation}
\tilde{u}_i(\nu)~=~\alpha_{ij}(\nu)u_j(\nu)+\beta_{ij}(\nu)u_j(\nu)^\dagger\,,
\label{eq:bogo-u-exact}
\end{equation}
using the Einstein summation convention, 
such that 
the coefficients $\alpha_{ij}$ and $\beta_{ij}$ depend on~$\mu$, 
and \eqref{eq:bogo-u-exact} reproduces \eqref{eqn:: first-order} to linear order in~$\mu$. 

We expand the Bogoliubov coefficients in \eqref{eq:bogo-u-exact} in $\mu$ as 
$\alpha_{ij}=\delta_{ij}+\mu\alpha^{(1)}_{ij} +O(\mu^2)$ and $\beta_{ij}=\mu\beta^{(1)}_{ij} +O(\mu^2)$. 
Consistency with \eqref{eqn:: first-order} implies 
\begin{equation}
\alpha^{(1)}=\begin{pmatrix}
	0 & 0 & \I \\
	0 & 0 & 0 \\
	\I & 0 & 0
	\end{pmatrix}
 \,, \quad\beta^{(1)}=\begin{pmatrix}
	0 & 0 & 0 \\
	0 & 0 & \I \\
	0 & \I & 0
	\end{pmatrix}\,, 
\end{equation}
where $\alpha^{(1)}_{XD}$ and 
$\beta^{(1)}_{YD}$ are read off directly from~\eqref{eqn:: first-order}, 
and $\alpha^{(1)}_{DX}$ and 
$\beta^{(1)}_{DY}$ are then determined by the Bogoliubov identities that follow from the preservation of the commutation relations~\cite{BD}. 
Writing 
$U= (u, u^\dagger)$, 
and similarly 
for~$\tilde{U}$, we then have 
\begin{align}
\tilde{U} = (I + \mu \gamma)U + O(\mu^2)
\,,
\label{eq:Utilde-linear}
\end{align}
where $U$ and $\tilde U$ are understood as column vectors, $I$ is the $6\times6$ identity matrix, and 
\begin{equation}
\gamma=\begin{pmatrix}
	\alpha^{(1)} & \beta^{(1)} \\
	\left(\beta^{(1)}\right)^* & \left(\alpha^{(1)}\right)^*
	\end{pmatrix}\,. 
\end{equation}

We promote the linear order Bogoliubov transformation \eqref{eq:Utilde-linear} into the nonperturbative Bogoliubov transformation $\tilde{U} = \exp(\mu \gamma)U$. Geometrically, $\exp(\mu \gamma)$ is the one-parameter subgroup of Bogoliubov transformations that is determined uniquely by the infinitesimal generator~$\gamma$, and it is thus the minimal nonperturbative completion of~\eqref{eq:Utilde-linear}. Further, as $\gamma^3=0$, the series for $\exp(\mu \gamma)$ terminates after the quadratic term: the geometric reason is that the Bogoliubov transformation matrices acting on $U$ are in the matrix group $U(3,3)$, and the subgroup $\exp(\mu \gamma)$ therein is of parabolic type, similar to null rotations in the Lorentz group. 

Writing out the series for the matrix exponential, we hence have 
\begin{equation}
\tilde{U}~= \left( I +\mu\gamma + \frac{1}{2}\mu^2\gamma^2 \right) \!  U\,,  
\label{eq:U-tilde-exact}
\end{equation}
as an exact relation. 
Writing \eqref{eq:U-tilde-exact} out in terms of $(X_\nu,Y_\nu,D_\nu)$, we then find 
\begin{subequations}
\label{eqn:: Bogo1XYD}
\begin{align}\label{eqn:: Bogo1X}
    \tilde{X}_\nu~=~&X_\nu \! \left(1-\frac{\mu^2}{2}\right)+\I\mu D_\nu - \frac{\mu^2}{2}Y_\nu^\dagger\,,\\
    \label{eqn:: Bogo1Y}
\tilde{Y}_\nu ~=~& Y_\nu \! \left(1+\frac{\mu^2}{2}\right)+\I\mu D_\nu^\dagger + \frac{\mu^2}{2}X_\nu^\dagger\,,\\
\label{eqn:: Bogo1D}
\tilde{D}_\nu ~=~& D_\nu+\I\mu \! \left(X_\nu+Y_\nu^\dagger\right)\,.
\end{align}\end{subequations} 
The quadratic terms in \eqref{eqn:: Bogo1XYD}
show that the interaction amplifies the laser modes~$Y_\nu$, de-amplifies the laser modes~$X_\nu$, and does neither to the BEC modes~$D_\nu$, despite the electromagnetic noise being injected into the BEC by the interaction. 

Finally, the post-interaction operators in \eqref{delta_atilde} and \eqref{eq:baa} in the main text are obtained from \eqref{eqn:: Bogo1X}
and 
\eqref{eqn:: Bogo1Y}
via 
\eqref{eqn:: inversi-common-and-difference-modes}
and 
\eqref{eq:XY-def} and their tilded counterparts.

\section{Backaction suppressed by squeezed laser fluctuations}

In this appendix we identify the mechanism by which squeezing the initial state of the laser beam fluctuations suppresses the backaction on the BEC\null. This mechanism leads to the suppressed added noise formula \eqref{diffPSDadd} in the main text, as we show in Supplemental Material~\cite{suppmat}. 
We follow the procedure introduced in~\cite{Unruh:2022gso}, adapting it to our notation.

In the notation of Appendix~B, we 
start with the pair of operators $(X_\nu,Y_\nu)$, where $0<\nu < \Delta$, and define the new pair $(\hat{X}_\nu,\hat{Y}_\nu)$ 
by the two-mode squeezing transformation 
\begin{subequations}
\label{eqn:: sq1and2}
\begin{align}
\label{eqn:: sq1}
\hat{X}_\nu~:=~&\cosh(\lambda)X_\nu+\sinh(\lambda)Y_\nu^\dagger\,,
\\
\label{eqn:: sq2}
\hat{Y}_\nu~:=~&\cosh(\lambda)Y_\nu+\sinh(\lambda)X_\nu^\dagger\,, 
\end{align}
\end{subequations}
where $\lambda\in\mathbb{R}$ is the squeezing parameter. 
The nonvanishing commutators of the new operators are 
$\bigl[ \hat{X}_\nu , \hat{X}_{\nu'} ^\dagger \bigr] = \bigl[ \hat{Y}_\nu , \hat{Y}_{\nu'} ^\dagger \bigr] = 2\pi \delta(\nu-\nu')$. The inverse transformation is 
\begin{subequations}
\label{eqn:: sqinv1and2}
\begin{align}
\label{eqn:: sqinv1}
X_\nu~=~&\cosh(\lambda)\hat{X}_\nu-\sinh(\lambda) \hat{Y}_\nu^\dagger\,,
\\
\label{eqn:: sqinv2}
Y_\nu~=~&\cosh(\lambda)\hat{Y}_\nu-\sinh(\lambda)\hat{X}_\nu^\dagger\,.   
\end{align}
\end{subequations}

The crucial observation is now that the post-interaction BEC annihilation 
operator 
$\tilde{D}_\nu$ \eqref{eqn:: Bogo1D} can be written in terms of the new operators $\hat{X}_\nu$ and $\hat{Y}_\nu$ as 
\begin{equation}
\tilde{D}_\nu 
~=~ 
D_\nu + \I\mu \E^{-\lambda}
\left(\hat{X}_\nu+\hat{Y}_\nu^\dagger\right)\,.
\end{equation}
This shows that if the laser beam fluctuations are initially prepared in the two-mode squeezed state $\ket{\lambda}$ that is annihilated by $\hat{X}_\nu$ and~$\hat{Y}_\nu$, 
the phonon number expectation value after the laser-BEC interaction is given by 
\begin{equation}\label{eqn:: BECnum}
   \braket{\tilde{D}_\nu^\dagger \tilde{D}_{\nu'}}_\lambda ~=~  \braket{D_\nu^\dagger D_{\nu'}}+\mu^2 \E^{-2\lambda} 2\pi\delta[\nu-\nu']\,,
\end{equation}
where the subscript $\lambda$ on the left-hand side denotes that the initial state of the laser fluctuations was~$\ket{\lambda}$. 
The $\mu^2$ term in \eqref{eqn:: BECnum} is the backaction noise, and the factor $\E^{-2\lambda}$ in this term shows that the noise is exponentially suppressed for large positive~$\lambda$. 

By Heisenberg's uncertainty principle, the exponential squeezing of $X_\nu+Y_\nu^\dagger = \E^{-\lambda}
(\hat{X}_\nu+\hat{Y}_\nu^\dagger)$
is concomitant with the exponential enhancement of the conjugate operator 
$-\I(X_\nu-Y_\nu^\dagger)=-\I\E^\lambda(\hat{X}_\nu-\hat{Y}_\nu^\dagger)$. 
Preparing the laser fluctuations in the state $\ket{\lambda}$ hence suppresses the BEC noise but enhances the conjugate electromagnetic noise. We shall show in Supplemental Material \cite{suppmat} that the balance of this suppression and enhancement leads to the total added noise formula \eqref{diffPSDadd} in the main text, allowing the total added noise to be below the standard quantum limit.

\clearpage \raggedbottom

\section{Supplemental Material}

Supplemental Material for 
``Nondestructive optomechanical detection scheme for Bose-Einstein condensates'' by 
Cisco Gooding,
Cameron R. D. Bunney,
Samin Tajik, 
Sebastian Erne,
Steffen Biermann,
J\"org Schmiedmayer,
Jorma Louko,
William G. Unruh and Silke Weinfurtner, 
doi.org/10.1103/4yfh-tm4f, published in 
Physical Review Letters (2026). 

\vspace{1ex}

\subsection{Heterodyne analysis: unsqueezed laser fluctuations}

In this section we give additional detail on the heterodyne detection scheme presented in the main text, establishing how the observed difference-photocurrent PSD $S_{ii}$ is related to the 
BEC PSD $S_{\phi\phi}$ by \eqref{diffPSDinf0} and \eqref{diffPSDadd0} when the laser fluctuations are initially in their vacuum state. 
The generalisation to a squeezed initial state for the laser fluctuations, leading to formula \eqref{diffPSDadd} in the main text, is considered below in Section~2.  

In the two-tone heterodyne scheme depicted in the optical circuit diagram of Figure~\ref{fig:exp}, 
the reference beam is generated by first applying a beamsplitter (BS) to the initial (modulated and filtered) laser; these beamsplitter outputs are the starting points for the signal and reference arms for the heterodyne scheme. An acousto-optic modulator (AOM) is applied to the reference beam, shifting the two modulation peaks in frequency by~$\Delta_{\mrm{LO}}$, such that $\Delta\ll\Delta_{\mrm{LO}}\ll2\Omega$. A phase-shifter (PS) is then used to give the frequency-shifted modulation bands a (tunable) relative phase~$\varphi$, such that the positive-frequency part of the reference beam takes the form
\begin{align}\label{LO}
   E_{\mrm{LO}}^+(t)~=~&E_0(\omega_0)\,|\beta|\,\E^{-\I(\omega_0 +\Delta_{\mrm{LO}})t}\nonumber\\
   &\times\frac{\left(\E^{-\I(\varphi+\Omega t)}-\E^{\I(\varphi+\Omega t)}\right)}{\sqrt{2}}
   \, .
\end{align}
The specific choice $\varphi=\psi_0$ --- in other words, tuning the relative phase to coincide with the constant bulk density shift from the BEC --- leads to a particularly appealing heterodyne signal, as we will see in \eqref{deltan0}. We assume $|\beta|\gg\alpha$ and treat the reference beam classically. 

The probe beam passes through the condensate (BEC). After the interaction, the positive-frequency part of the fluctuations in the probe field are given by $\E^{-\I \omega_0 t}(\E^{-\I(\Omega t+\psi_0)}\delta \tilde{a}_+(t)+\E^{\I(\Omega t+\psi_0)}\delta \tilde{a}_-(t))$. The probe field is combined with the reference field determined by \eqref{LO} using a beamsplitter, producing output fields $(\tilde{E}(t)\pm E_{\mrm{LO}}(t))/\sqrt{2}$. Defining the photon flux operator as the product of the negative-frequency and positive-frequency parts of the electric field, the photon fluxes of the beamsplitter outputs are identified as $(\tilde{E}^-(t)\pm E_{\mrm{LO}}^-(t))(\tilde{E}^+(t)\pm E_{\mrm{LO}}^+(t))/2$. These two photon fluxes are converted to photocurrents by photodiodes, and then subtracted from each other, producing the \textit{difference photocurrent} $i(t)$; this constitutes our heterodyne signal. Conceptually, the difference photocurrent can be interpreted as the outcome of a continuous measurement of the difference of two photon flux operators, one for each detected beamsplitter output. We denote this \textit{difference photon flux} operator by $n(t)$.

In the context of analysing measurement noise inherent to our detection scheme, the fluctuation operator $\delta n(t)\equiv n(t)-\langle n(t)\rangle$ is more informative than the difference photon flux operator itself, as it directly quantifies deviations from the mean signal, that is, the noise, which is the primary quantity of interest in our analysis.
Moreover, the form of the reference field in~\eqref{LO} was chosen to allow the detection scheme to operate in the vicinity of a ``dark port'' setting: in the intermediate-frequency (IF) regime determined by $\Delta_{\mrm{LO}}$, cancellations occur for ``bright'' (common-mode) contributions to the difference photocurrent, greatly improving contrast for measuring relatively ``dark'' (difference-mode) contributions relating to our BEC signal. The same logic forms the basis for `balanced' detection schemes, whereby an expectation value of an individual operator is no more than a means to an end. In what follows, we will neglect the expected difference photocurrent, and focus entirely on its fluctuations, $\delta n(t)=n(t)-\langle n(t)\rangle$.

Our analysis of the resulting photocurrent power spectral density (PSD) will parallel the heterodyne treatment of Bowen and Milburn~\cite{bowenmilburn}. The basis of this treatment is Glauber's theory of photodetection~\cite{glauber63}, which describes the photocurrent PSD as a result of two-time photon coincidences, expressed in terms of normally-ordered creation and annihilation operators. The treatment is then specialised to linear detection of optical fields, in which case Glauber's expression for the photocurrent PSD $S_{ii}[\omega]$ coincides with the frequency-symmetrised PSD of the detected photon flux, $\bar{S}_{nn}[\omega]=\frac{1}{2}\left(S_{nn}[\omega]+S_{nn}[-\omega]\right)$, where $S_{nn}[\omega]$ is the (unsymmetrised) photon flux PSD. Explicitly, we have  
\begin{align}
 S_{ii}[\omega]\approx \bar{S}_{nn}[\omega]\equiv \frac{1}{2}\left(S_{nn}[\omega]+S_{nn}[-\omega]\right)\, .
\end{align}
For nonstationary signals, the photon flux PSD $S_{nn}[\omega]$ can be calculated using the general expression for cross-spectral densities characterising power associated with noise correlations between two given operators $A$ and~$B$,
\begin{equation}\label{eq:iPSD}
S_{AB}[\omega
]=\lim_{T\rightarrow\infty}\frac{1}{T}\iint_{-T/2}^{T/2}\D t\,\D t'\,\E^{-\I\omega (t-t')}\langle \delta A(t)^\dagger \delta B(t')\rangle\,,
\end{equation}
in terms of the noise operators $\delta A=A-\langle A\rangle$ (and similarly for $B$). Hence, the difference photon flux PSD $S_{nn}[\omega]$ is defined by setting $A=B=n$ in~\eqref{eq:iPSD}. 
For a thorough discussion of technical subtleties associated with the relation between photocurrent PSDs and electromagnetic correlators, see~\cite{Ou1987}. 

We now calculate the PSD for the difference photocurrent generated in our heterodyne scheme, which corresponds to a detection of the difference photon flux $\delta n(t)$. In this case, it is straightforward to show that the difference photon flux PSD is symmetric in frequency, i.e. $\bar{S}_{nn}[\omega]=S_{nn}[\omega]$ \cite{bowenmilburn}; consequently, the difference photocurrent PSD can be calculated using the simple relation
\begin{align}
    S_{ii}[\omega]\approx S_{nn}[\omega]\, .
\label{eq:Sii-approx-Snn}
\end{align}
To obtain the explicit form of $S_{nn}[\omega]$, we use the general cross-spectral density definition \eqref{eq:iPSD} with $A=B=n$. Using the notation $\tilde{Z}(t)=(\delta\tilde{a}_+(t)-\delta\tilde{a}_-(t))/\sqrt{2}$ for the post-interaction difference-mode operator in the time domain and setting $\varphi=\psi_0$ for the rest of the analysis, we can express the difference photon flux as 
\begin{align}\label{deltandecomp}
    \delta n(t)=\delta n_0(t)+\Delta_0(t)+\Delta_0^\dagger(t)\, ,
\end{align}
with the definitions
\begin{align}\label{deltan0}
    \delta n_0(t)=|\beta|\left(\E^{\I\Delta_\mrm{LO}t}\tilde{Z}(t)+\E^{-\I\Delta_\mrm{LO}t}\tilde{Z}^\dagger(t)\right)
\end{align}
and
\begin{align}
    &\Delta_0(t)=\\
    &\frac{|\beta|}{\sqrt{2}}\E^{\I\Delta_\mrm{LO}t}\left(\E^{2\I(\Omega t+\psi_0)}\delta \tilde{a}_-(t)-\E^{-2\I(\Omega t+\psi_0)}\delta \tilde{a}_+(t)\right)\, .\nonumber
\end{align}
The choice $\varphi=\psi_0$ is directly responsible for the cancellation of phases in \eqref{deltan0} that results in the appearance of $\tilde{Z}$ and $\tilde{Z}^\dagger$ operators, rather than other combinations of $\delta \tilde{a}_\pm$ and $\delta \tilde{a}_\pm^\dagger$ operators.

Based on the role played by the unequal-time correlator in the integrand of~\eqref{eq:iPSD} and the PSD identification~\eqref{eq:Sii-approx-Snn}, it is clear that the difference photocurrent PSD we seek to calculate depends exclusively on unequal-time photon flux correlations. Inserting the photon flux decomposition~\eqref{deltandecomp} into $\langle \delta n(t)^\dagger \delta n(t')\rangle$, we identify three distinct types of terms. First, we have $\langle \delta n_0(t)^\dagger \delta n_0(t')\rangle$; this term produces contributions to the photon flux PSD of the form
\begin{align}
   S_{n_0 n_0}[\omega]=|\beta|^2\left( S_{ \tilde{Z}  \tilde{Z}}[\Delta_{\mrm{LO}}+\omega]+S_{ \tilde{Z}^\dagger  \tilde{Z}^\dagger}[\Delta_{\mrm{LO}}-\omega]\right)\, .
\end{align}

Second, we find unequal time correlations of $\Delta_0(t)+\Delta_0^\dagger(t)$. Of the four correlators appearing upon expansion of the two sums, the two terms with phase factors containing $t+t'$ will produce rapid oscillations of the integrand in the $T\rightarrow\infty$ limit, contributing negligibly to the overall PSD; neglecting such contributions is known as the rotating-wave approximation. Only the other two terms, with phase factors containing $t-t'$, contribute to the desired PSD:
\begin{align}
    \langle \Delta_0(t)\Delta_0^\dagger(t')\rangle +\langle \Delta_0^\dagger(t)\Delta_0(t')\rangle\, .
\end{align}
When evaluating each of these correlators, we will again apply the rotating-wave approximation, but this time to terms with phase factors $\sim \E^{\pm2\I\Omega(t+t')}$, since the modulation frequency is still much greater than the BEC frequencies of interest; as a result, we find that $\langle \Delta_0(t)\Delta_0^\dagger(t')\rangle$ is $|\beta|^2 \E^{\I\Delta_\mrm{LO}(t-t')}/2$ multiplied by
\begin{align}
    \E^{-2\I\Omega(t-t')}\langle \delta \tilde{a}_+(t) \delta \tilde{a}_+^\dagger(t')\rangle+\E^{2\I\Omega(t-t')}\langle \delta \tilde{a}_-(t) \delta \tilde{a}_-^\dagger(t')\rangle
\end{align}
and $\langle \Delta_0^\dagger(t)\Delta_0(t')\rangle$ is $|\beta|^2 \E^{-\I\Delta_\mrm{LO}(t-t')}/2$ multiplied by
\begin{align}
    \E^{2\I\Omega(t-t')}\langle \delta \tilde{a}_+^\dagger(t) \delta \tilde{a}_+(t')\rangle+\E^{-2\I\Omega(t-t')}\langle \delta \tilde{a}_-^\dagger(t) \delta \tilde{a}_-(t')\rangle\, .
\end{align}
These correlators therefore contribute to the photon flux PSD as $|\beta|^2/2$ multiplied by
\begin{align}
    &S_{ \tilde{a}_+  \tilde{a}_+}[\Delta_{\mrm{LO}}+\omega-2\Omega]+ S_{ \tilde{a}_-  \tilde{a}_-}[\Delta_{\mrm{LO}}+\omega+2\Omega]\nonumber\\
    +&S_{ \tilde{a}_+^\dagger  \tilde{a}_+^\dagger}[\Delta_{\mrm{LO}}-\omega-2\Omega]+S_{ \tilde{a}_-^\dagger  \tilde{a}_-^\dagger}[\Delta_{\mrm{LO}}-\omega+2\Omega]\, .
\end{align}

Finally, we find cross terms that have correlations between $\delta n_0$ and either $\Delta_0$ or $\Delta_0^\dagger$ - these terms will have isolated factors of either $\E^{\pm 2\I \Omega t}$ or $\E^{\pm 2\I \Omega t'}$; consequently, the rapid oscillations will vanish upon integration, as per the rotating-wave approximation. 
Combining these results, we obtain
\begin{widetext}
\begin{align}
    \frac{S_{nn}[\omega]}{|\beta|^2}=&S_{ \tilde{Z}  \tilde{Z}}[\Delta_{\mrm{LO}}+\omega]+S_{ \tilde{Z}^\dagger  \tilde{Z}^\dagger}[\Delta_{\mrm{LO}}-\omega]\\
    &+\frac{1}{2}\left(S_{ \tilde{a}_+  \tilde{a}_+}[\Delta_{\mrm{LO}}+\omega-2\Omega]+ S_{ \tilde{a}_-  \tilde{a}_-}[\Delta_{\mrm{LO}}+\omega+2\Omega]+S_{ \tilde{a}_+^\dagger  \tilde{a}_+^\dagger}[\Delta_{\mrm{LO}}-\omega-2\Omega]+S_{ \tilde{a}_-^\dagger  \tilde{a}_-^\dagger}[\Delta_{\mrm{LO}}-\omega+2\Omega]\right)\, .\nonumber
\end{align}
Normalising by $|\beta|^2$ and evaluating at $\omega=\Delta_{\mrm{LO}}-\nu$, the photocurrent PSD is found to be
\begin{align}
\label{PSDsqueezed0}
    S_{ii}[\Delta_\mrm{LO}-\nu]=&S_{ \tilde{Z}  \tilde{Z}}[2\Delta_{\mrm{LO}}-\nu]+S_{ \tilde{Z}^\dagger  \tilde{Z}^\dagger}[\nu]\\
    &+\frac{1}{2}\left(S_{ \tilde{a}_+  \tilde{a}_+}[2(\Delta_{\mrm{LO}}-\Omega)-\nu]+ S_{ \tilde{a}_-  \tilde{a}_-}[2(\Delta_{\mrm{LO}}+\Omega)-\nu]+S_{ \tilde{a}_+^\dagger  \tilde{a}_+^\dagger}[\nu-2\Omega]+S_{ \tilde{a}_-^\dagger  \tilde{a}_-^\dagger}[\nu+2\Omega]\right)\, .\nonumber
\end{align}
\end{widetext}

Let us now work out each term in~\eqref{PSDsqueezed0} explicitly. Recall from 
\eqref{delta_atilde} and \eqref{eq:baa} in the main text that 
\begin{equation}
\label{eq:atildeplus-suppmat}
\delta\tilde{a}_\pm[\nu]~=~\delta a_\pm[\nu]\pm\frac{\I \mu D_\nu}{\sqrt{2}} \pm \frac{\mu^2}{4} \sgn(\nu)\delta a_{b}[\nu]\,,
\end{equation}
where 
\begin{equation}
    \delta a_{b}[\nu]=\delta a_-[\nu]+\delta a_-[-\nu]^\dagger-\delta a_+[\nu]-\delta a_+[-\nu]^\dagger\, .
\end{equation} 
The time-domain version of \eqref{eq:atildeplus-suppmat} is 
\begin{equation}\label{atildepm}
\delta\tilde{a}_\pm(t)~=~\delta a_\pm(t)\pm\frac{\I \mu \phi(t)}{\sqrt{2}} \pm \Delta_b(t)\,,
\end{equation}
where the backaction operator $\Delta_b(t)$ is defined as
\begin{equation}
\Delta_b(t)=\frac{\mu^2}{4}\int \frac{d\nu}{2\pi} \E^{-\I\nu t}\text{sgn}(\nu)\delta a_b[\nu]\,.
\label{eq:Deltabt}
\end{equation}
An expression for $\tilde{Z}(t)$ analogous to ~\eqref{atildepm} is given by
\begin{equation}\label{Ztilde}
\tilde{Z}(t)~=~Z(t)+\I \mu \phi(t)+\sqrt{2}\,\Delta_b(t)\,.
\end{equation}
The terms in~\eqref{PSDsqueezed0} can then be calculated using the definition~\eqref{eq:iPSD} by working out the unequal-time correlators for the various operators, and performing the time integrals in the $T\rightarrow\infty$ limit. 

Keeping in mind that 
$S_{\phi\phi}[2\Delta_{\mrm{LO}}-\nu]$ can be neglected because 
$S_{\phi\phi}[\nu]$ is nonvanishing only for $-\Delta < \nu < \Delta$ whereas $\Delta \ll \Delta_{\mrm{LO}}$, the contributions to ~\eqref{PSDsqueezed0} from $S_{\tilde{Z}\tilde{Z}}$ and $S_{\tilde{Z}^\dagger\tilde{Z}^\dagger}$ can be expressed as
\begin{align}\label{StildZtildZ}
    S_{\tilde{Z}\tilde{Z}}[2\Delta_{\mrm{LO}}-\nu]=2S_{\Delta_b \Delta_b}[2\Delta_{\mrm{LO}}-\nu]=\frac{\mu^4}{8}
\end{align}
and
\begin{align}\label{SZdagZdag}
    S_{\tilde{Z}^\dagger \tilde{Z}^\dagger}[\nu]=\frac{1}{2}+\mu^2 S_{\phi\phi}[\nu]+\frac{\mu^4}{8}+\sqrt{2}(S_{Z^\dagger \Delta_b^\dagger}+S_{\Delta_b^\dagger Z^\dagger})[\nu]\, ,
\end{align}
where the shorthand $(S_{ab}+S_{cd})[\nu]=S_{ab}[\nu]+S_{cd}[\nu]$ was used, along with the backaction-backaction correlators
\begin{align}
    \langle \Delta_b(t)^\dagger \Delta_b(t')\rangle=\langle \Delta_b(t) \Delta_b(t')^\dagger\rangle=\frac{\mu^4}{16}\delta(t-t')
\end{align}
that follow from~\eqref{eq:Deltabt}.

The remaining imprecision-backaction cross spectra $S_{Z^\dagger \Delta_b^\dagger}[\nu]$ and $S_{\Delta_b^\dagger Z^\dagger}[\nu]$ in ~\eqref{SZdagZdag} can be calculated using the correlator
\begin{align}
    &\langle \delta a_\pm(t) \Delta_b(t')^\dagger\rangle
    \nonumber\\
    &=\mp\frac{\mu^2}{4}\int \frac{\D\nu'}{2\pi} \E^{\I \nu' t'}\text{sgn}(\nu')
    \langle \delta a_\pm(t)\delta a_\pm[\nu']^\dagger\rangle\, ,
\end{align}
along with its complex conjugate. To evaluate this, we consider the mixed time-frequency correlator, 
\begin{align}
    \langle \delta a_\pm(t)\delta a_\pm[\nu']^\dagger\rangle=\int \frac{\D\nu}{2\pi} \E^{-\I\nu t}\langle\delta a_\pm[\nu]\delta a_\pm[\nu']^\dagger\rangle=\E^{-\I\nu' t}\, .
\end{align}
Hence,
\begin{align}
    \langle \delta a_\pm(t) \Delta_b(t')^\dagger\rangle
     =
    \mp\frac{\mu^2}{4} 
    \int \frac{\D\nu'}{2\pi} \E^{-\I \nu'(t-t')}\text{sgn}(\nu') \,.
\label{imp-back}
\end{align}
The integral in \eqref{imp-back} evaluates to a multiple of the principal value of $1/(t-t')$, but it is convenient to leave the integral unevaluated for the moment. Inserting the result \eqref{imp-back} into $S_{a_+^\dagger \Delta_b^\dagger}[\nu]$, we can then interchange the order of integration between the time and frequency integrals, in which case the time integration produces $\delta(\nu+\nu')$. Performing the remaining (trivial) frequency integration and repeating the same steps for the conjugate correlator $\langle \Delta_b(t)\delta a_\pm(t')^\dagger\rangle$, we find
\begin{align}\label{aDelta_cross}
    S_{a_\pm^\dagger \Delta_b^\dagger}[\nu]=S_{\Delta_b^\dagger a_\pm^\dagger}[\nu]=\pm\frac{\mu^2}{4}\text{sgn}(\nu)\, .
\end{align}
It follows that the desired imprecision-backaction cross spectra are given by
\begin{align}\label{ZDelta_cross}
    S_{Z^\dagger \Delta_b^\dagger}[\nu]=S_{\Delta_b^\dagger Z^\dagger}[\nu]=\frac{\mu^2}{2\sqrt{2}}\text{sgn}(\nu)\, .
\end{align}

Next we find the contributions to 
$S_{ii}[\Delta_{\mrm{LO}}-\nu]$ from $S_{ \tilde{a}_+  \tilde{a}_+}$ and $S_{ \tilde{a}_-  \tilde{a}_-}$ to be
\begin{align}
    S_{ \tilde{a}_+  \tilde{a}_+}[2(\Delta_{\mrm{LO}}-\Omega)-\nu]=S_{\Delta_b \Delta_b}[2(\Delta_{\mrm{LO}}-\Omega)-\nu]=\frac{\mu^4}{16}
\end{align}
and
\begin{align}
     S_{ \tilde{a}_-  \tilde{a}_-}[2(\Delta_{\mrm{LO}}+\Omega)-\nu]=S_{\Delta_b \Delta_b}[2(\Delta_{\mrm{LO}}+\Omega)-\nu]=\frac{\mu^4}{16}\, .
\end{align}
The remaining terms can be expressed as
\begin{align}\label{adagadagplus}        S_{\tilde{a}_+^\dagger \tilde{a}_+^\dagger}[\nu-2\Omega]=
\frac{1}{2}+(S_{a_+^\dagger \Delta_b^\dagger}+S_{\Delta_b^\dagger a_+^\dagger})[\nu-2\Omega]+\frac{\mu^4}{16}
\end{align}
and
\begin{align} \label{adagadagminus}      S_{\tilde{a}_-^\dagger \tilde{a}_-^\dagger}[\nu+2\Omega]=
\frac{1}{2}-(S_{a_-^\dagger \Delta_b^\dagger}+S_{\Delta_b^\dagger a_-^\dagger})[\nu+2\Omega]+\frac{\mu^4}{16}\, ,
\end{align}
which include vacuum noise terms $S_{a_\pm^\dagger a_\pm^\dagger}[\nu\mp2\Omega]=1/2$. The final imprecision-backaction cross spectra left to evaluate satisfy
\begin{align}\label{crossPSD1}
    S_{a_+^\dagger \Delta_b^\dagger}[\nu-2\Omega]=S_{\Delta_b^\dagger a_+^\dagger}[\nu-2\Omega]
\end{align}
and
\begin{align}\label{crossPSD2}
    S_{a_-^\dagger \Delta_b^\dagger}[\nu+2\Omega]=S_{\Delta_b^\dagger a_-^\dagger}[\nu+2\Omega]\, .
\end{align}
However, applying the same procedure that was used to derive ~\eqref{aDelta_cross} and ~\eqref{ZDelta_cross} produces $\delta(\nu\pm 2\Omega+\nu')$ factors, which vanish upon integration over the frequency band since $|\nu'|\ll |\nu\pm 2\Omega|$.

Collecting results, we arrive at
\begin{align}\label{twotonePSD0}
    S_{ii}[\Delta_\mrm{LO}-\nu]=&1+\mu^2 \left(S_{\phi\phi}[\nu]+\text{sgn}(\nu)\right)+\frac{3\mu^4}{8}\, ,
\end{align}
which is the content of formulas \eqref{diffPSDinf0}
and \eqref{diffPSDadd0} in the main text.

\subsection{Heterodyne analysis: squeezed laser fluctuations}

In this section we generalise the difference-photocurrent PSD \eqref{twotonePSD0} to the case when the laser fluctuations are initially in the squeezed state defined in Appendix~C\null, establishing the added noise formula \eqref{diffPSDadd} in the main text. 

In the notation of Appendices B and~C, we work with the operators $(X_\nu,Y_\nu)$ and $(\hat{X}_\nu,\hat{Y}_\nu)$, 
related by~\eqref{eqn:: sq1and2}, 
where $0<\nu < \Delta$, and 
the parameter $\lambda\in\mathbb{R}$ in \eqref{eqn:: sq1and2} is the squeezing parameter. We assume that the laser fluctuations are prepared in the initial state $\ket{\lambda}$ that is annihilated by $\hat{X}_\nu$ and~$\hat{Y}_\nu$. 
Using \eqref{eqn:: sqinv1and2} and a subscript $\lambda$ to indicate that the initial state of the laser fluctuations is~$\ket{\lambda}$, we then find the expected initial number operators
\begin{align}\label{sqcorr1}
\braket{X_\nu^\dagger X_{\nu'}}_\lambda  
=\braket{Y_\nu^\dagger Y_{\nu'}}_\lambda=
\sinh^2 \! \lambda\, 2\pi\delta[\nu-\nu']\, ,
\end{align}
from which we can use the commutation relations $[X_\nu,X_{\nu'}^\dagger]=[Y_\nu,Y_{\nu'}^\dagger]=2\pi \delta[\nu-\nu']$ to obtain
\begin{align}\label{sqcorr2}
\braket{X_\nu X_{\nu'}^\dagger}_\lambda  
=\braket{Y_\nu Y_{\nu'}^\dagger}_\lambda=
\cosh^2 \! \lambda\, 2\pi\delta[\nu-\nu']\, ,
\end{align}
There are also initial cross-correlators, given by
\begin{align}\label{sqcorr3}
\langle X_\nu Y_{\nu'}\rangle_\lambda 
&= \langle Y_\nu X_{\nu'}\rangle_\lambda= \langle X_\nu^\dagger Y_{\nu'}^\dagger\rangle_\lambda=\langle Y_\nu^\dagger X_{\nu'}^\dagger\rangle_\lambda\,
\notag\\
&= 
-\sinh\lambda\, \cosh\lambda\, 2\pi\delta[\nu-\nu'] \,,
\end{align}
with all other combinations vanishing.

With the above results, we can now recall the relations 
$\delta \tilde{a}_\pm[\nu]=(1/\sqrt{2})(z_\nu\pm\tilde{X}_\nu)$ for $\nu>0$ and $\delta \tilde{a}_\pm[\nu]=(1/\sqrt{2})(z_\nu\pm\tilde{Y}_{-\nu})$ for $\nu<0$ to express each relevant term of ~\eqref{PSDsqueezed0} in terms of $X$ and $Y$ operators, and repeat the analysis of the previous section. For this calculation, it is convenient to write $Z(t)=(\delta a_+(t)-\delta a_-(t))/\sqrt{2}$ as
\begin{align}\label{Zspectral}
    Z(t)~=~\int_0^\Delta \frac{\D\nu}{2\pi}\left(\E^{-\I\nu t}X_\nu+\E^{\I\nu t}Y_\nu\right)\, .
\end{align}
Then, introducing the notation $\Delta A_\nu \equiv \tilde{A}_\nu-A_\nu$ for $A\in\{X,Y,D,Z\}$, the Bogoliubov transformation \eqref{eqn:: Bogo1XYD} can be re-expressed as
\begin{subequations}
\begin{align}
    \Delta X_\nu~=~&\tilde{X}_\nu-X_\nu=\I\mu D_\nu-\frac{\mu^2}{2}(X_\nu+Y_\nu^\dagger)\,,\\
\Delta Y_\nu ~=~&\tilde{Y}_\nu-Y_\nu= \I\mu D_\nu^\dagger+\frac{\mu^2}{2}(Y_\nu+X_\nu^\dagger)\,,\\
\Delta D_\nu ~=~&\tilde{D}_\nu-D_\nu= \I\mu(X_\nu+Y_\nu^\dagger)=\I\mu \E^{-\lambda}(\hat{X}_\nu+\hat{Y}_\nu^\dagger)\, .
\end{align}\end{subequations} 
In the time domain, we will also write
\begin{align}\label{DeltaZ}
    \Delta Z(t)=\tilde{Z}(t)-Z(t)=\int_0^\Delta \frac{\D\nu}{2\pi}\left(\E^{-\I\nu t}\Delta X_\nu+\E^{\I\nu t}\Delta Y_\nu\right)
\end{align}
and, to abuse notation with $\Delta a_\pm(t)\equiv \delta \tilde{a}_\pm(t)-\delta a_\pm(t)$, 
\begin{align}\label{Deltaa}
    \Delta a_\pm(t)&=\pm\frac{1}{\sqrt{2}}\Delta Z(t)= \pm\left( \frac{\I\mu}{\sqrt{2}} \phi(t)+\Delta_b(t)\right)\, ,
\end{align}
in terms of the backaction operator $\Delta_b(t)$ introduced in \eqref{eq:Deltabt}.

The contribution from $S_{\tilde{Z}\tilde{Z}}$ is obtained from the correlator $\langle \tilde{Z}(t)^\dagger\tilde{Z}(t')\rangle_\lambda$, which can be decomposed into four terms via $\tilde{Z}(t)=Z(t)+\Delta Z(t)$. Each of the four resulting terms can then be evaluated with judicious use of the spectral representations \eqref{eq:Deltabt} and \eqref{DeltaZ}, along with the squeezed-state correlations \eqref{sqcorr1}-\eqref{sqcorr3}. This leads to the PSD
\begin{align}
    S_{\tilde{Z}\tilde{Z}}[2\Delta_\mrm{LO}-\nu]=\frac{\sinh^2 \! \lambda}{2}+\frac{\mu^4}{8}\E^{-2\lambda}\, . 
\end{align}

Similarly, $S_{\tilde{Z}^\dagger \tilde{Z}^\dagger}[\nu]$ is determined by the correlator $\langle \tilde{Z}(t)\tilde{Z}(t')^\dagger\rangle_\lambda$, though in this case there are more subcomponents to evaluate. Specifically, there are nonzero contributions given by
\begin{align}
    S_{\tilde{Z}^\dagger \tilde{Z}^\dagger}[\nu]=&S_{Z^\dagger Z^\dagger}[\nu]+\mu^2 S_{\phi\phi}[\nu]+2S_{\Delta_b^\dagger \Delta_b^\dagger}[\nu]\nonumber\\
    &+\sqrt{2}\left(S_{Z^\dagger \Delta_b^\dagger}[\nu]+S_{\Delta_b^\dagger Z^\dagger}[\nu]\right)\, .
\end{align}
First, we calculate $S_{Z^\dagger Z^\dagger}$, which yields
\begin{align}
 S_{Z^\dagger Z^\dagger}[\nu]=\frac{\cosh^2 \! \lambda}{2}\, .  
\end{align}
The backaction-backaction contribution is
\begin{align}
 S_{\Delta_b^\dagger \Delta_b^\dagger}[\nu]=S_{\Delta_b\Delta_b}[2\Delta_\mrm{LO}-\nu]=\frac{\mu^4}{16}\E^{-2\lambda}\, ,
\end{align}
and the cross-correlations take the form
\begin{align}
 S_{ Z^\dagger \Delta_b^\dagger}[\nu]=S_{\Delta_b^\dagger Z^\dagger}[\nu]=\frac{\mu^2}{2\sqrt{2}}\cosh\lambda\,\E^{-\lambda}\text{sgn}(\nu)\, ,
\end{align}
which implies
\begin{align}
    &S_{\tilde{Z}^\dagger\tilde{Z}^\dagger}[\nu]=\\
    &\frac{\cosh^2 \! \lambda}{2}+\mu^2\left(S_{\phi\phi}[\nu]+\E^{-\lambda}\cosh\lambda\,\text{sgn}(\nu)\right)+\frac{\mu^4}{8}\E^{-2\lambda}\, .\nonumber
\end{align}

The remaining contributions are from $S_{\tilde{a}_\pm\tilde{a}_\pm}$ and $S_{\tilde{a}_\pm^\dagger\tilde{a}_\pm^\dagger}$, which we decompose further using \eqref{Deltaa}. The terms without a $\Delta$ represent shot noise; for these, we will need the correlators
\begin{align}
    \langle \delta a_\pm^\dagger(t)\delta a_\pm(t')\rangle_\lambda=\frac{1}{2}\langle Z(t)^\dagger Z(t')\rangle_\lambda
\end{align}
and
\begin{align}
    \langle \delta a_\pm(t)\delta a_\pm^\dagger(t')\rangle_\lambda&=\frac{1}{2}\left(\langle z(t) z^\dagger(t')\rangle_\lambda+\langle Z(t) Z^\dagger(t')\rangle_\lambda\right)\nonumber\\
    &=\frac{1}{2}\left(\frac{1}{2}\delta(t-t')+\langle Z(t) Z^\dagger(t')\rangle_\lambda\right)\, .
\end{align}
One then finds
\begin{align}
    S_{a_\pm a_\pm}=\frac{1}{2}S_{ZZ}=\frac{\sinh^2 \! \lambda}{4}
\end{align}
and
\begin{align}
    S_{a_\pm^\dagger a_\pm^\dagger}=\frac{1}{4}+\frac{1}{2}S_{Z^\dagger Z^\dagger}=\frac{1+\cosh^2 \! \lambda}{4}\, .
\end{align}

Of the terms with $\Delta$ appearing twice, the contributions from $S_{\phi\phi}$ evaluated far outside its domain of support vanish, as do the cross spectra between $\phi$ and $\Delta_b$. The only nonvanishing terms arise from backaction-backaction correlations; hence, we have
\begin{align}
S_{\Delta a_\pm\Delta a_\pm}=S_{\Delta_b\Delta_b}=\frac{\mu^4}{16}\E^{-2\lambda}
\end{align}
and
\begin{align}
S_{\Delta a_\pm^\dagger\Delta a_\pm^\dagger}=S_{\Delta_b^\dagger\Delta_b^\dagger}=\frac{\mu^4}{16}\E^{-2\lambda}\, .
\end{align}

Finally, the terms with $\Delta$ appearing only once can only be purely electromagnetic, due to the vanishing of the one-point function for $\phi$. These terms therefore represent cross spectra between shot noise and backaction. In the unsqueezed case, these terms cancel in pairs; in the squeezed case, however, there are twice as many terms to consider, since $S_{a_\pm \Delta_b}\neq 0$. Half the terms, then, are of the new form,
\begin{widetext}
\begin{align}\label{firstlast}
    S_{ a_+\Delta_b}[2(\Delta_{\mrm{LO}}-\Omega)-\nu]+S_{ \Delta_b a_+}[2(\Delta_{\mrm{LO}}-\Omega)-\nu]- S_{ a_-\Delta_b}[2(\Delta_{\mrm{LO}}+\Omega)-\nu]-S_{\Delta_b a_-}[2(\Delta_{\mrm{LO}}+\Omega)-\nu]\, .
\end{align}
Using the relations $S_{a_\pm\Delta_b}=\pm\frac{1}{\sqrt{2}} S_{Z\Delta_b}=\pm\frac{1}{\sqrt{2}} S_{\Delta_b Z}=S_{\Delta_b a_\pm}$,
we find that  \eqref{firstlast} reduces to 
\begin{align}
    \sqrt{2}\left(S_{ Z\Delta_b}[2(\Delta_{\mrm{LO}}-\Omega)-\nu]+S_{ Z\Delta_b}[2(\Delta_{\mrm{LO}}+\Omega)-\nu]\right)\, .
\end{align}
These terms vanish for the same reason as ~\eqref{crossPSD1} and ~\eqref{crossPSD2}, though they would formally cancel even if the frequency band was infinitely extended: if we follow the steps that led to ~\eqref{aDelta_cross} and ~\eqref{ZDelta_cross} and naively carry out the frequency integration over the whole real line, the result is proportional to $\text{sgn}(2(\Delta_{\mrm{LO}}-\Omega)-\nu)+\text{sgn}(2(\Delta_{\mrm{LO}}+\Omega)-\nu)$. Since we are assuming $\Omega\gg\Delta_{\mrm{LO}}$, the terms cancel. 

The other half are the terms \eqref{crossPSD1}, \eqref{crossPSD2} that vanished in the unsqueezed case:
\begin{align}\label{secondlast}
    S_{ a_+^\dagger\Delta_b^\dagger}[\nu-2\Omega]+S_{ \Delta_b^\dagger a_+^\dagger}[\nu-2\Omega]-S_{ a_-^\dagger\Delta_b^\dagger}[\nu+2\Omega]-S_{ \Delta_b^\dagger a_-^\dagger}[\nu+2\Omega]\, .
\end{align}
Considerations analogous to those applied to \eqref{firstlast} lead to the conclusion that the terms in \eqref{secondlast} vanish in the squeezed case as well.

Altogether, we are left with
\begin{align}\label{twotonePSDsqueeze}
    S_{ii}[\Delta_\mrm{LO}-\nu]=1+\frac{3}{2}\sinh^2 \! \lambda+\mu^2 \left(S_{\phi\phi}[\nu]+\E^{-\lambda}\cosh\lambda\,\text{sgn}(\nu)\right)+\frac{3\mu^4}{8}\E^{-2\lambda}\, ,
\end{align}
which is the content of formulas \eqref{diffPSDinf0}
and \eqref{diffPSDadd} in the main text.
\end{widetext}

\subsection{Comparative analysis of incident power threshold and the standard quantum limit}
In this section, we will explore the ratio between the laser probe power threshold and the power required to reach the SQL. As mentioned in the Letter, the SQL is reached by tuning the coherent amplitude $\alpha$ of the laser probe to $\alpha_{\text{SQL}}^2=4 \sqrt{2}/(\sqrt{3}\varepsilon^2)$. The corresponding laser probe power will be denoted by $P_\mrm{SQL}$; when the laser probe power reaches this value, the measurement noise reaches a minimum (in the absence of correlations between shot noise and backaction). Hence, the ratio we consider in this section determines the feasibility of attaining the SQL: if we can manage to arrange a unit ratio ($\bar{P}/P_\mrm{SQL}=1$) experimentally while maintaining a reasonable threshold on the power entering the BEC, then the SQL can be reached through nondestructive continuous measurement; otherwise, we cannot. We will work in natural units until restoring SI units in the final ratios.

To begin, we note that the laser power $P_0$ for each individual  modulation band is $P_0= \omega_0 \alpha^2$, and the total laser probe power averaged over the modulation cycles is $\bar{P}=2 P_0=2\omega_0 \alpha^2$. In terms of the total laser power (or, equivalently, the coherent amplitude~$\alpha$), 
the photon scattering rate 
$\Gamma_{\text{sc}}$ is
\begin{equation}\label{scatrat}
    \Gamma_{\mathrm{sc}}=\frac{4\hat{\alpha}_I  \bar{P}}{\pi r_0^2}=\frac{8\hat{\alpha}_I  \omega_0 \alpha^2}{\pi r_0^2}.
\end{equation}
Here, $\omega_r=\omega_0$, is the resonance frequency, and $\delta_0$ is the detuning in units of the half-linewidth. Using this, the total laser power is:
\begin{equation}
    \bar{P}=\frac{\Gamma_{\mathrm{sc}} \pi r_0^2}{4\hat{\alpha}_I}\, .
\end{equation}

Now, the power required to reach the SQL ($\mu^2=\mu^2_{SQL}=2\sqrt{2/3}$) can be expressed as
\begin{equation}
P_{\text{SQL}}= 2\omega_0 \alpha_{\text{SQL}}^2=\frac{8 \sqrt{2}\omega_0}{\varepsilon^2\sqrt{3}}\, ,
\end{equation}
similarly averaged over modulation cycles. The polarisability~\eqref{eqn:: polarizability} also obeys $\hat{\alpha}_I=-\hat{\alpha}_R/\delta_0$ and $\hat{\alpha}_R\delta_0=-24\pi^2/\omega_0^3$, from which it follows that the ratio of the total laser power to the power needed to reach the SQL (reinstating SI units) is given by
\begin{equation}\label{PowRat}
\frac{\bar{P}}{P_{\text{SQL}}}=\frac{3\sqrt{3}\Gamma_{sc}  \pi^3 r_0^2 m \rho_0 c^2} { \sqrt{2} \,\omega_0^2\hbar  }\, .
\end{equation}
For the system parameters considered in ~\cite{PhysRevLett.125.213603} ($r_0=3\mu\mrm{m}$, $\rho_0=10^3\mu\mrm{m}^{-2}$, $m=133\mrm{amu}$, and $\omega_0/(2\pi)=10^{14}\mrm{Hz}$, summarised in Table~\ref{Table}), we find the estimate
\begin{equation}
\frac{\bar{P}}{P_{\text{SQL}}}\approx 492\cdot \Gamma_{sc}/\text{Hz}\, ,
\end{equation}
indicating that the SQL can be reached while upholding a scattering rate threshold of $\Gamma_{sc}\approx 0.00203\text{Hz}$, well within bounds for maintaining a nondestructive BEC measurement.

For the squeezed states discussed above, we define $P_\lambda=2\omega_0\alpha_{\lambda}^2$ as the laser power that minimizes the added noise for our squeezed states, with $\alpha_\lambda$ being the corresponding coherent amplitude, derived from $\mu_\lambda$ \eqref{mulambda}. The power ratio \eqref{PowRat} then generalises to
\begin{equation}\label{PowRatsqueeze}
\frac{\bar{P}}{P_\lambda}=\frac{3\sqrt{3}\Gamma_{sc}  \pi^3 r_0^2 m \rho_0 c^2 }{\sqrt{2} \E^\lambda \sqrt{1+\frac{3}{2}\sinh^2 \! \lambda}\, \omega_0^2\hbar  }\, .
\end{equation}
From the analysis in the main text, the added noise is minimal for $\lambda = \frac12\ln\bigl(5+4\sqrt{10}\bigr)-\ln3 \approx 0.3367$; in this case, the system parameters given above imply
\begin{equation}
\frac{\bar{P}}{P_\lambda}\approx 324\cdot \Gamma_{sc}/\text{Hz}\, ,
\end{equation}
indicating that the SQL can be optimally beaten while maintaining a scattering rate threshold of $\Gamma_{sc}\approx 0.00309\text{Hz}$, again consistent with a nondestructive measurement.

The scattering rates calculated here correspond to timescales $1/\Gamma_{sc}$. These are considerably longer than typical timescales of both BEC dynamics and BEC lifetimes. The operating conditions for sub-SQL sensitivity in our scheme are therefore compatible with detection of modes within the phononic band. Due to the weak dependence of the ratios \eqref{PowRat} and \eqref{PowRatsqueeze} on~$\rho_0$, our conclusions here are robust to a wide range of BEC densities.

\begin{table}[t!]
\centering
\caption{BEC experimental parameters~\cite{PhysRevLett.125.213603}.}
\begin{tabular}{|c|c|} 
 \hline
 \textbf{Parameter} & \textbf{Value} \\ 
 \hline
 $r_0$ & $3~\mu\mrm{m}$ \\ 
 $\rho_0$ & $10^3~\mu\mrm{m}^{-2}$ \\  
 $m$ & $133~\mrm{amu}$ \\   
 $\omega_0/2\pi$ & $10^{14}~\mrm{Hz}$ \\  
 \hline
\end{tabular}
\label{Table}
\end{table}

\subsection{$\mu$-sensitivity of added noise for $\nu>0$ versus $\nu<0$}

Figure \ref{fig:SQL} of the main text shows that added noise is much more sensitive to the intensity parameter $\mu$ in the $\nu<0$ regime than in the $\nu>0$ regime. In this section we show that the physical mechanism responsible for this asymmetry is the same as in backaction cooling in the measurement of zero point mechanical oscillations in a cavity optomechanical system~\cite{Khalili2012}.

In~\cite{Khalili2012}, a mechanical oscillator is coupled to two optical modes: a strong red-detuned ``cooling'' (pump) mode and a weak ``readout'' (probe) mode. When the pump is tuned below cavity resonance by one mechanical frequency (the anti-Stokes condition), radiation-pressure fluctuations remove phonons from the oscillator to cool it down to its ground state. This backaction cooling arises because quantum fluctuations of the radiation-pressure force scale with the measurement strength, and the corresponding terms in the PSD scale as the square of this measurement strength. It is shown in \cite{Khalili2012} that the measured spectrum takes the form 
\begin{align}
S_{yy}(\omega)
=\frac{S_{zz}}{\alpha^2}
+2 \Realpart\left[
\chi^*(\omega),S_{zF}(\omega)
\right]
+\alpha^2 S_{FF}^{\rm BA}(\omega)
+S_{xx}^{q},
\label{eq:suppmat-posnegnu-analogue}
\end{align}
where the $1/\alpha^2$ term is the shot-noise contribution, the $\alpha^2$ term is the backaction force~PSD, and the $\alpha$-independent cross-correlation term encodes the Stokes/anti-Stokes asymmetry. 

Now, formula \eqref{eq:suppmat-posnegnu-analogue} is directly analogous to our noise expression~\eqref{diffPSDadd}, 
\begin{equation}
\label{eq:sm-diffPSDadd}
    \mathcal{N}[\nu;\mu,\lambda]=\frac{1+\frac{3}{2}\sinh^2 \! \lambda}{\mu^2}+\frac{3\mu^2\, \E^{-2\lambda}}{8}+\E^{-\lambda}\cosh\lambda\,\text{sgn}(\nu)\,,
\end{equation}
where our intensity parameter $\mu$ is the analogue of~$\alpha$. 
Our $\nu<0$ branch hence corresponds to the anti-Stokes (cooling) sideband, which is a backaction dominated regime. In this branch, the $\mu^2$ backaction term interferes with the $\nu<0$ offset, producing strong sensitivity to~$\mu$. Our $\nu>0$ branch, by contrast, corresponds to the Stokes (excitation) sideband, where the $\nu>0$ offset suppresses the interference, the noise floor is dominated by the shot noise term $1/\mu^2$, and the overall response varies only weakly with~$\mu$. 

\clearpage

\end{appendix}

\end{document}